\documentclass[fleqn,usenatbib,a4paper]{article}

\let\counterwithin\relax
\usepackage{chngcntr}
\usepackage[T1]{fontenc}
\usepackage[utf8]{inputenc}
\usepackage[english]{babel}
\usepackage{newtxtext,newtxmath}
\usepackage[title,titletoc,toc]{appendix}
\usepackage{graphics,float}
\usepackage{amsmath}	% Advanced maths commands
\usepackage{subfigure}
\usepackage{graphicx}
\usepackage{float}
\usepackage[colorinlistoftodos]{todonotes}
\usepackage[colorlinks=true, allcolors=blue]{hyperref}
\usepackage{verbatim}
\usepackage[authoryear,round,longnamesfirst]{natbib}
\usepackage{chngcntr}
\usepackage[labelfont=bf]{caption}
\usepackage{authblk}

\usepackage[font=small,format=hang,labelfont={sf,bf}]{caption}
\usepackage[a4paper,top=3cm,bottom=2cm,left=3cm,right=3cm,marginparwidth=1.75cm]{geometry}

\defcitealias{tv1}{Golkhou \& Butler 2014}
\defcitealias{tv2}{Golkhou et al. 2015}
\defcitealias{tv3}{Sonbas et al. 2015}
\defcitealias{lu}{L\"u et al. 2012}
\defcitealias{neucosma1}{H\"ummer et al. 2010}
\defcitealias{neucosma2}{H\"ummer et al. 2012}

\title{Constraining the contribution of Gamma-Ray Bursts to the high-energy diffuse neutrino flux with 10 years of ANTARES data}

\author[1,2]{A.~Albert}
\author[3]{M.~Andr\'e}
\author[4]{M.~Anghinolfi}
\author[5]{G.~Anton}
\author[6]{M.~Ardid}
\author[7]{J.-J.~Aubert}
\author[8]{J.~Aublin}
\author[8]{B.~Baret}
\author[9]{S.~Basa}
\author[10]{B.~Belhorma}
% ...
\author[ ]{(ANTARES Collaboration)}

\affil[1]{\scriptsize{Universit\'e de Strasbourg, CNRS,  IPHC UMR 7178, F-67000 Strasbourg, France}}
\affil[2]{\scriptsize Universit\'e de Haute Alsace, F-68200 Mulhouse, France}
\affil[3]{\scriptsize{Technical University of Catalonia, Laboratory of Applied Bioacoustics, Rambla Exposici\'o, 08800 Vilanova i la Geltr\'u, Barcelona, Spain}}
\affil[4]{\scriptsize{INFN - Sezione di Genova, Via Dodecaneso 33, 16146 Genova, Italy}}
\affil[5]{\scriptsize{Friedrich-Alexander-Universit\"at Erlangen-N\"urnberg, Erlangen Centre for Astroparticle Physics, Erwin-Rommel-Str. 1, 91058 Eabrlangen, Germany}}
\affil[6]{\scriptsize{Institut d'Investigaci\'o per a la Gesti\'o Integrada de les Zones Costaneres (IGIC) - Universitat Polit\`ecnica de Val\`encia. C/  Paranimf 1, 46730 Gandia, Spain}}
\affil[7]{\scriptsize{Aix Marseille Univ, CNRS/IN2P3, CPPM, Marseille, France}}
\affil[8]{\scriptsize{Universit\'e de Paris, CNRS, Astroparticule et Cosmologie, F-75006 Paris, France}}
\affil[9]{\scriptsize{Aix Marseille Univ, CNRS, CNES, LAM, Marseille, France }}
\affil[10]{\scriptsize{National Center for Energy Sciences and Nuclear Techniques, B.P.1382, R. P.10001 Rabat, Morocco}}
% ...
\begin{document} % End of preamble and beginning of text.
\maketitle % Produces the title.
\begin{abstract}
\noindent
Addressing the origin of the astrophysical neutrino flux observed by IceCube is of paramount importance.  Gamma-Ray Bursts (GRBs) are among the few astrophysical sources capable of achieving the required energy to contribute to such neutrino flux through p$\gamma$ interactions. In this work, ANTARES data have been used to search for upward going muon neutrinos in spatial and temporal coincidence with 784 GRBs occurred from 2007 to 2017. For each GRB, the expected neutrino flux has been calculated in the framework of the internal shock model and the impact of the lack of knowledge on the majority of source redshifts and on other intrinsic parameters of the emission mechanism has been quantified. It is found that the model parameters that set the radial distance where shock collisions occur have the largest impact on neutrino flux expectations. In particular, the bulk Lorentz factor of the source ejecta and the minimum variability timescale are found to contribute significantly to the GRB-neutrino flux uncertainty. For the selected sources, ANTARES data have been analysed, by maximising the discovery probability of the stacking sample through an extended maximum-likelihood strategy. Since no neutrino event passed the quality cuts set by the optimisation procedure, 90 \% confidence level upper limits (with their uncertainty) on the total expected diffuse neutrino flux have been derived, according to the model. The GRB contribution to the observed diffuse astrophysical neutrino flux around 100~TeV is constrained to be less than 10 \%.
\end{abstract}

\section{Introduction}
High-energy astrophysical neutrinos were discovered few years ago \citep{icecube1,icecube2,icecube3}, opening a new window to the study of the Universe. Identifying the sources of these neutrinos is one of the key scientific targets of the astroparticle physics community. The most powerful accelerators are needed to explain the energetics of these neutrinos and it is possible that their sources generate also Ultra-High-Energy Cosmic-Rays (UHECRs), the most energetic particles observed to date, with energies above $10^9$~GeV \citep{nu_crs1,nu_crs2,nu_crs3,uhecrs1,uhecrs2,uhecrs3, globus}. Therefore, the discovery of neutrino sources might guide us towards the solution of the one-century-old mystery about the origin of such charged particles.\\
Among several astrophysical sources, Gamma-Ray Bursts (GRBs) are considered one of the most promising candidate sources of astrophysical neutrinos. They are intense flashes of high-energy electromagnetic radiation, observed isotropically in the sky \citep{grb1}, and thus believed to be of extragalactic nature. GRBs constitute the most powerful known explosions in the Universe, releasing energies between $10^{51}$ and $10^{54}$~ergs in a few seconds. For detailed reviews about GRBs see \citet{piran}, \citet{meszaros} and \citet{zhang}.\\\\GRBs have historically been observed by space-based facilities, through photons in the energy band from the keV to hundreds of GeV \citep{energy_grb}. Recently, the first detections of photons in the sub-TeV energy band from GRB180720B \citep{hessTeVgrb}, GRB190114C \citep{magicTeVgrb} and from the low-luminous GRB190829A \citep{hessTeVgrb2,hessTeVgrb2_2} have been carried out with ground-based imaging atmospheric Cherenkov telescopes. Such a novel energetic component has provided further evidence of the powerfulness of this class of accelerators. However, all these sub-TeV observations are thought to be related to the afterglow component of the emission, that is expected when the jet impinges upon the circumstellar medium \citep{sahu2020,hessTeVgrb2_afterglow}. On the other hand, the prompt component, that should be produced within the region of particle acceleration, has not been observed yet in TeV gamma rays. The lack of prompt TeV gamma rays seems to be mostly connected to the difficulty faced by ground-based telescopes to follow-up the GRB event within the few seconds of duration of the prompt phase. Nonetheless, the discovery of TeV emission has renewed the discussion about the hadronic versus leptonic origin of the observed radiation. Though leptonic scenarios are typically favoured in GRB modelling, the highest energy photons might be witnesses of the onset of a hadronic component \citep{ghisellini2020}. This fact has clear implications in a multi-messenger scenario, from the point of view of both follow-up and offline analysis of coincident high-energy neutrinos (see e.g. \citet{antaresICRC2019}).\\Multi-messenger searches targeted at GRBs appear very promising; being transients and extremely energetic explosions, these sources allow to strongly reduce the background during their very short duration. If hadrons are accelerated in GRBs, neutrinos are expected to be produced by the collisions of protons (or heavier nuclei) on the intense radiation field of the jet. Neutrinos are ideal messengers in the search for distant astrophysical objects, being electrically neutral, stable and weakly interacting particles. Thus, unlike protons or charged nuclei, neutrinos are not diverted in their path from their source to the Earth. In addition, unlike photons, neutrinos are not absorbed while propagating towards the Earth. For these reasons, searching for a temporal and spatial coincidence among GRB photons and high-energy neutrinos is crucial to safely identify this kind of sources as hadronic factories and, in addition, to shed light on the composition of their jets.\\\\Over the past years, the two major neutrino telescopes of the Northern and Southern hemispheres, respectively ANTARES \citep{antares} and IceCube \citep{icecube}, have been searching for neutrino signals coincident with GRBs in time and direction. The lack of detections from these searches has allowed to set progressively stronger upper limits, thus limiting also the possible contribution of these sources to the observed astrophysical diffuse neutrino flux. Nonetheless, current limits do not yet provide significant constraints on the validity of the internal shock model \citep{is}, once the many uncertainties on parameters that affect the predictions are taken into account.
\\The results of previous searches of high energy neutrinos emitted by GRBs with ANTARES data can be found in \citet{antares_grb1,antares_grb2,antares_bright}, while for IceCube in \citet{icecube_grb3, icecube_grb1, icecube_grb2}. In the present paper, the search for astrophysical neutrinos from GRBs is extended, including almost 10 years of ANTARES data. This work differs from previously published results, since it focuses on improving the predictions on the expected neutrino fluences from GRBs. This is achieved by considering the wealth of information accumulated so far thanks to the many astronomical observations, rather than assuming some fixed standard values that do not correctly reproduce the properties of the source sample. Contextually, the different uncertainties due to the poor knowledge of the source dynamics are taken into account and propagated on the produced neutrino spectrum, with the aim of providing a clear understanding of the assumptions and limitations behind the upper limits that are set.\\\\ 
The paper is structured as follows. In Sec.~\ref{sec:antares_detector} the ANTARES detector and the data acquisition system are described. In Sec.~\ref{sec:parameter_selection} the adopted sample for this analysis and the criteria used for selecting the GRB parameters are explained. In Sec.~\ref{sec:model} the neutrino spectra predicted by the internal shock model are discussed, focusing on the uncertainties due to the poor knowledge of some parameters. In Sec.~\ref{sec:detector} the analysis chain is described, explaining the Monte Carlo (MC) simulations of GRB neutrino events that provide the detector response to the signal. Then, in Sec.~\ref{sec:background} the estimation of the background that characterises ANTARES data is presented. In Sec.~\ref{sec:pseudo-exp} the analysis optimisation is discussed, through the set up of MC pseudo-experiments generated with the aim of obtaining the highest discovery potential for the neutrino flux, by exploiting an extended maximum-likelihood ratio statistical method.
In Sec.~\ref{sec:optimisation} the diffuse search performed through the stacking technique, investigating whether the discovery potential can be improved by limiting the analysis to an optimised sub-sample of bursts, is presented. Finally, in Sec.~\ref{sec:results} and Sec.~\ref{sec:conclusions} the results of our analysis are shown.

\section{ANTARES detector and data taking}\label{sec:antares_detector}
ANTARES \citep{antares} is a large volume water-Cherenkov neutrino telescope in the Northern hemisphere, located in the deep water of the Mediterranean Sea, offshore Toulon (France), and fully operational since May 2008. Due to its performances and characteristics, the detector is primarily sensitive to neutrinos in the TeV-PeV energy range. The most relevant neutrino signals for the study of astrophysical sources are the track-like signatures provided by muons, produced by $\nu_{\mu}$ charged-current interactions. In this channel, about 50 \% of the track events are reconstructed within 0.4$^{\circ}$ of the parent neutrino for an $E^{-2}$ neutrino spectrum \citep{giulia_illuminati}. The remaining interaction channels produce hadronic and electromagnetic showers, that are observed inside the detector as spheres of light radially propagating from the interaction vertex and whose direction is reconstructed with an angular uncertainty of few degrees. For this reason, in this work the analysis is focused on the track-like signals with better angular resolution.\\
From the experimental point of view, track-like signals can either be the real tracks induced by muons or misidentified showers (incorrectly reconstructed as tracks). In order to take into account this possibility, the electron neutrino interactions are also simulated and the track-like events reconstructed from this channel are included in the analysis. In order to reduce the very abundant background coming from atmospheric muons, only upward going events are considered. However, given the very high statistics of atmospheric muons, these have to be further reduced by cuts on the track reconstruction quality. This selection leaves an irreducible background made of atmospheric neutrinos \citep{gaisser}.

\begin{table}
	\centering
	\caption{Percentage contributions of the different satellite catalogues to the determination of GRB position and spectrum. The position of the burst is taken from the detector with the smallest angular error. The spectrum is taken from the satellite with the most extended energy band. The total sample is made up of 784 GRBs.}
	\label{tab:statistics1}
	\begin{tabular}{lcc} 
		\hline
		\hline
		Source & Position & Spectrum\\
		\hline
		Swift & 29.9\% & 16.7\%\\
		~~~~Swift-BAT & 9.3\% & \\
		~~~~Swift-UVOT & 3.4\% & \\
		~~~~Swift-XRT & 17.2\% & \\
		Fermi & 68.8\% & 71.6\% \\
		other (e.g. Konus-Wind) & 1.3\% & 11.7\%\\
		\hline
	\end{tabular}
\end{table}
\begin{figure}
\centering
	\includegraphics[width=0.7\columnwidth]{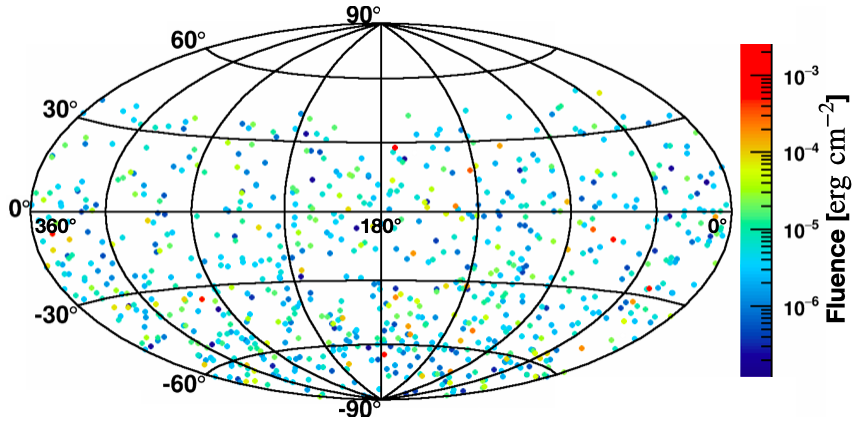}
    \caption{Sky distribution and fluence of the selected 784 GRBs in equatorial coordinates.}
    \label{fig:skymap}
\end{figure}

\section{GRB selection and parameters}
\label{sec:parameter_selection}
The GRB parameters needed for the search (time, direction) and the simulation of expected neutrino fluxes (photon spectrum, fluence, redshift) are collected from published results of Swift\footnote{Swift catalogue in \url{https://swift.gsfc.nasa.gov/archive/grb_table/}}\citep{swift}, Fermi\footnote{Fermi-GBM in \url{https://heasarc.gsfc.nasa.gov/W3Browse/fermi/fermigbrst.html} \citep{gbm1,gbm2,gbm3}. Fermi-LAT in \citet{fermilat_catalogue}.} \citep{lat,gbm} and Konus-Wind\footnote{Konus-Wind information is only available through the GCN archive: \url{http://gcn.gsfc.nasa.gov/gcn3_archive.html}} \citep{kw1}. Starting with a full sample of GRBs that includes 2604 sources, a selection is performed, satisfying the following criteria: \\
(i) Short burst are excluded, as this class is poorly understood in terms of neutrino production during their short prompt phase. In other words, only GRBs with prompt duration\footnote{T$_{90}$ is the time in which 90 \% of the gamma-ray fluence is emitted, during the so-called prompt phase.} T$_{90} \geq 2$~s (the so-called long GRBs) are selected. \\
(ii) Coordinates of the bursts should be meaured by at least one satellite. Those GRBs such that the angular uncertainty provided by the satellite is larger than 10$^{\circ}$ are excluded.\\
(iii) The gamma-ray spectrum has to be measured.
This is typically fitted with a broken power-law, a cut-off power-law or a smoothly broken power-law function. It is also required that the spectral indices satisfy the conditions $\gamma_{1}>-4$ and $\gamma_{2}>-5$, where $\gamma_1$ and $\gamma_2$ are respectively the slope below and above the energy break. \\
(iv) At least one parameter among electromagnetic fluence and redshift has to be measured, since their values are needed in the calculation of the source luminosity, that is primarily affecting the yields in both gamma rays and neutrinos. \\
(v) Only GRBs that were below the ANTARES horizon at trigger time have been selected.\\
When physical parameters of a GRB are measured by different detectors, the adopted criteria are:\\
(i) The burst's position is taken from the detector with the smallest angular error (typically Swift-UVOT, then Swift-XRT, Fermi-LAT, Swift-BAT and finally Fermi-GBM).\\
(ii) The burst's duration, spectrum and fluence are taken from the satellite reporting measurements in the most extended energy band (typically Konus-Wind $0.02-10$~MeV, then Fermi $0.01-1$~MeV, and finally Swift $0.015-0.15$~MeV).\\\\
Following these criteria, 488 more GRBs have been added with respect to the ones analysed in \citet{antares_grb1}. The final sample contains 784 GRBs and their spatial distribution in the equatorial sky is shown in Fig.~\ref{fig:skymap}. The field of view of the ANTARES detector for upward going events is $2\pi$ sr and, due to its geographical location, the sky up to a declination of $47^\circ$ is visible. The statistics of parameters adopted in this analysis from the several instruments about the source positioning and spectral modeling is specified in Tab.~\ref{tab:statistics1}. Note that in some cases some parameters have not been measured, e.g., in many cases the information on the energy break is missing, as well as the spectral slope above it.
In such a situation, default values are assumed: the peak energy of the burst is set at 200~keV when unknown (33 \% of the cases) and $\gamma_2=\gamma_1-1$ when only $\gamma_1$ is available from catalogs (1.4 \% of the cases). Moreover, the host galaxy of the GRB can fail to be identified by the multi-wavelength follow-up and so the redshift remains unknown. With respect to the redshift, former analyses have been adopting the default value $z=2.15$ in case this information was not available.
In addition, for the minimum variability timescale $t_{\rm{v}}$ of the bursts, which can be determined by the width of the peaks in the light curve, a default value of $t_{\rm{v}}=10$~ms (derived from theoretical consideration put forward in \citet{guetta}), has been used so far in all neutrino searches. However, since these parameters affect crucially the GRB-neutrino fluence estimation, a different strategy has been here adopted, as explained in Sec.~\ref{uncertainties}.

\begin{figure}
\centering
	\includegraphics[width=0.7\columnwidth]{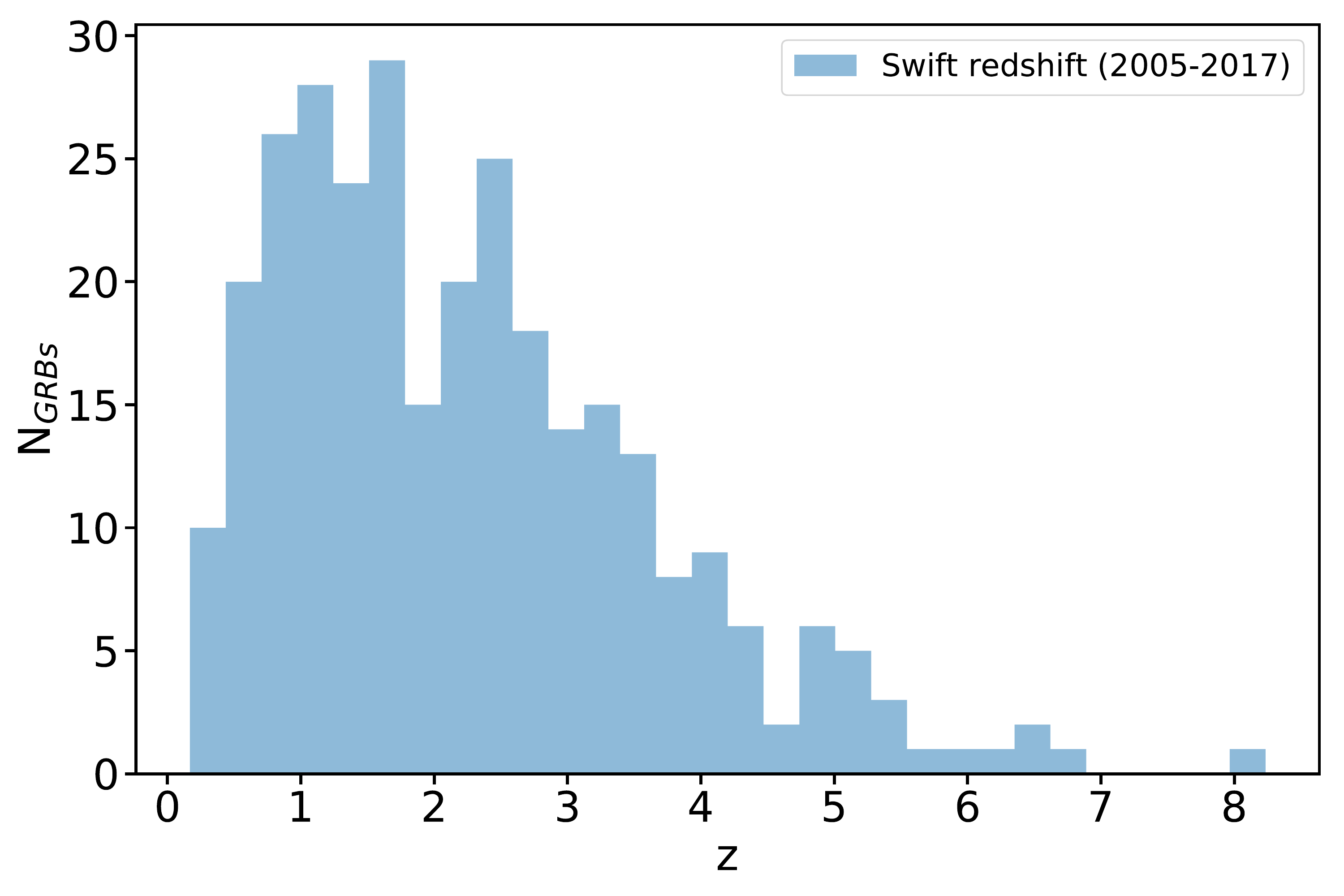}
    \caption{Swift redshift distribution for GRBs detected from 2005 to 2017 (data are available in \url{https://swift.gsfc.nasa.gov/archive/grb_table/}).}
    \label{fig:swift_distribution}
\end{figure}

\section{Computation of the neutrino flux from internal shocks}\label{sec:model}
The most commonly accepted scenario to explain the physics of GRBs is the so-called fireball model \citep{fireball}, where the stellar explosion drives the relativistic expansion of a plasma of particles. According to the internal shock framework of the fireball model \citep{is2,is3,is}, the central engine of GRBs produces multiple shells with different speeds: the faster ones catch up with the slower ones and collide. The acceleration mechanism converts part of the jet's kinetic energy into internal energy \citep{piran} and a fraction of this energy is expected to be transferred to non-thermal particles, achieving relativistic speeds. Accelerated electrons subsequently loose their energy through synchrotron and inverse Compton processes. The intense emitted radiation field constitutes the target for photo-hadronic interactions with the protons accelerated at shock fronts: from these collisions, mesons are produced, which then decay, generating neutrinos and gamma rays. These processes constitute the so-called prompt phase of the emission. Nonetheless, if GRBs were purely leptonic sources \citep{leptonic}, the observed radiation would be completely ascribed to processes involving primary electrons, such that there would be no possibility to produce neutrinos in these sources.\\
In a simplified one-zone emission model, a single representative collision is realized at the so-called internal shock radius, located at a distance 
\begin{equation}
    \mathrm{R_{is}}\simeq \frac{2\Gamma^2 c t_{\rm v}}{(1+z)} \simeq 2 \times 10^{13} \left( \frac{t_v}{0.01~{\rm s}} \right) \left( \frac{\Gamma}{10^{2.5}} \right)^2 \left( \frac{3}{1+z} \right) \, {\rm cm}.
    \label{eq:is}
\end{equation} 
from the central emitter. Note that the internal shock radius strongly affects the characteristic energy range of emitted neutrinos, while simultaneously scaling the normalization of the neutrino spectrum \citep{guetta}.
As Eq.~\eqref{eq:is} shows, the Lorentz factor impacts significantly the spectral modeling. In addition, the variability time $t_v$ is expected to be a crucial parameter as well, given its broad range of variation among GRBs. It is also worth mentioning that some models \citep{lyutikov,kumar2008} have argued emission radii larger than what indicated by Eq.~\eqref{eq:is}, correspondingly predicting a less efficient neutrino production. Interestingly, these models favor the interpretation of GRBs are sources of UHECRs \citep{murase2,he2012}, as heavy nuclei would be allowed to survive without being disintegrated.\\\\
Furthermore, neutrino production is thought to be efficiently realized also at radii below the photosphere, namely the location where the optical depth of Thomson scattering along the jet falls to unity, which is expected to be located at $\rm R_{ph} \sim 10^{12}$~cm. In the photospheric scenario \citep{ph1,ph2,ph3,tv_th3,ph4,ph5}, because the dissipation radius is located closer to the central engine ($\rm R_{ph}<R_{is}$), the characteristic energy range where photospheric neutrinos are expected to be detected is typically lower than what is expected in the internal shock model. It follows that, in order to test the photospheric model, special data acquisition conditions are required so as to access events with a low level trigger. The interested reader is referred to \citet{antares_bright} for a dedicated study on the photospheric model as applied to some interesting bright GRB events. In turn, the present work will be focused on testing the internal shock scenario.\\\\
The neutrino flux expected from GRBs during the prompt phase was first computed analytically by \citet{first_calculation} and \citet{wb1997}, while refined calculations were performed in the following years (\citet{guetta,murase,murase2,neucosma1,neucosma2}). Among such approaches, the numerical method developed by \citet{neucosma1} and, later on, by \citetalias{neucosma2} is adopted in the present work.

\begin{figure}
    \centering
	\includegraphics[width=0.7\columnwidth]{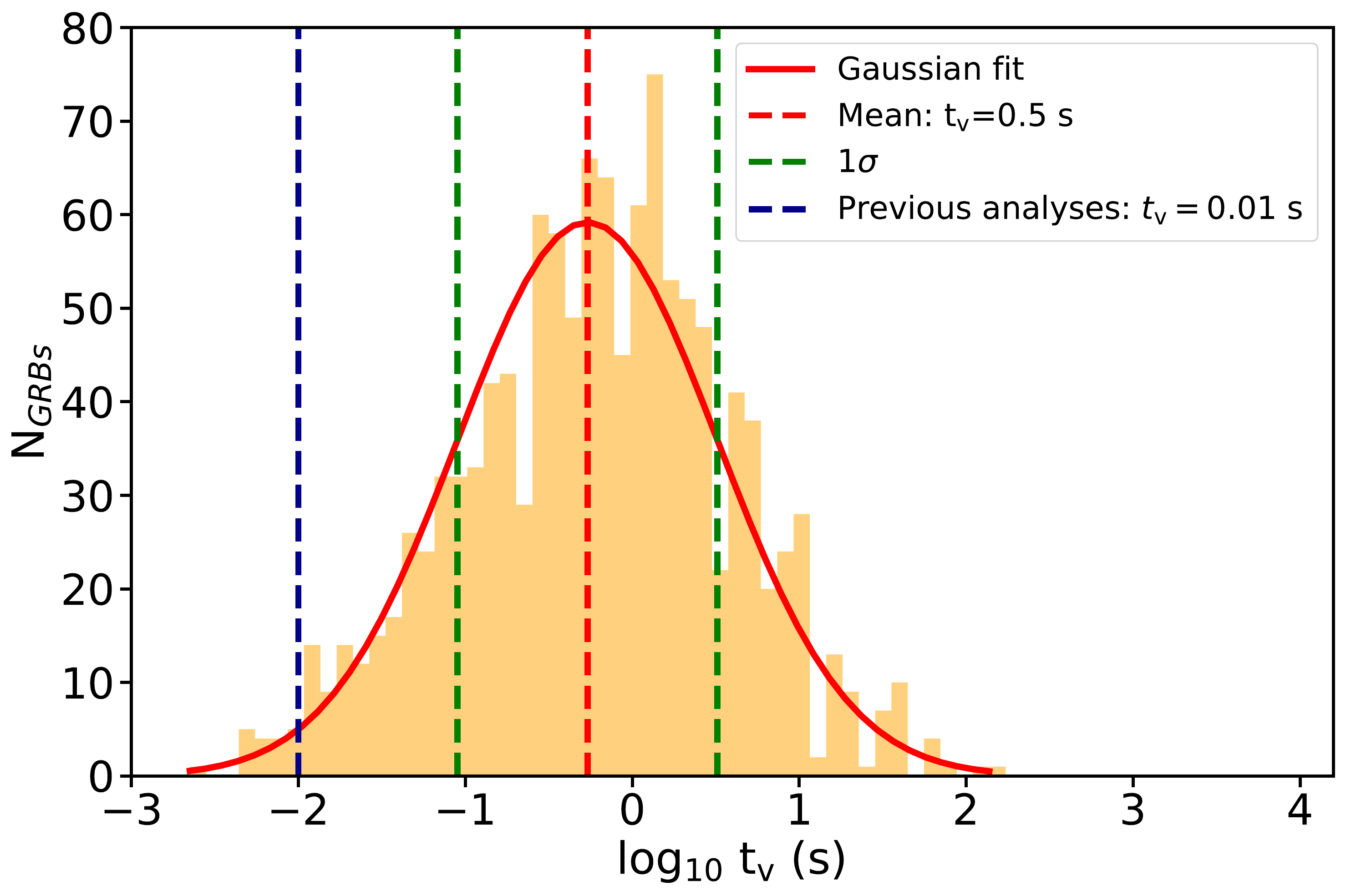}
    \caption{Distribution of minimum variability timescales obtained analysing 1213 GRB light curves \citepalias{tv1,tv2,tv3}. The solid red line indicates the Gaussian fit of the distribution. The dashed red line is the mean of the distribution, from which a mean value of $t_{\mathrm{v}}=0.5$ s is obtained. The dashed green lines indicate the $1\sigma$ level. In dashed blue the default value previously used in \cite{antares_grb1}, $t_{\rm v}=10$~ms, is indicated.}
    \label{fig:tv_distribution}
\end{figure}

\subsection{The numerical modelling with NeuCosmA}
\label{neucosma}
The event generator \textquoteleft Neutrinos from Cosmic Accelerator' (NeuCosmA) \citepalias{neucosma1,neucosma2}, used in this work to compute the expected neutrino fluxes, is based on the assumption that protons are accelerated through first-order Fermi processes \citep{firstorderFermi} (i.e. with a differential energy spectrum $\propto E^{-2}$) in the relativistic ejecta of the burst and interact with the intense jet photon field. The latter is described by an energy distribution in the form of a broken power-law function \citep{band}, constrained by observations.\\
The adopted version of NeuCosmA assumes a one-zone collision, namely it simulates average shell properties, such as an average shock speed or Lorentz factor $\Gamma$ (i.e. the bulk Lorentz factor of the jet). Indeed, it can be considered as approximation that the ejecta coast with constant bulk $\Gamma$ before decelerating due to the interaction with the external medium \citep{zhang}. Note that in a more realistic situation, the collisions between plasma shells are different one from the other, each happening under different physical conditions, as the irregular burst light curves demonstrate. The latest release of the NeuCosmA code allows to account for such a multi-collisions scenario \citep{multicollision,neucosma_multicollision}, by modelling the specific light curve of individual GRBs. However, given the extended sample of sources considered in this work, the one-zone collision approach, that rather relies on the average spectral properties of the bursts, is adopted.\\\\\\
Since the synchrotron-emitted photons constitute the radiation field on which accelerated protons collide, the normalisation of the neutrino fluence depends linearly on the intensity of the photon flux and on the ratio of fireball energy in protons to electrons. This so-called baryonic loading, $f_{\rm p}$, is an unknown of the problem, possibly constrained by neutrino observations. From the theoretical point of view, a reasonable value for it could be $f_{\rm p}\simeq 10$ \citepalias{neucosma2}; such a value will be fixed in the following for each GRB considered. The normalisation of the neutrino fluence depends on other several quantities \citepalias{neucosma2}:\\
(i) The total fraction of the energy transferred from protons to pions. Considering the reaction kinematics, approximately 20 \% of the proton energy is transferred to the produced pion in each interaction.\\
(ii) The isotropic gamma-ray luminosity of the burst, L$_{\gamma,\mathrm{iso}}$. It is given by L$_{\gamma,\mathrm{iso}}=4 \pi \rm \rm d_{L}^2F_{\gamma}/\mathrm{\mathrm{T_{90}}}$, where F$_{\gamma}$ is the bolometric gamma-ray fluence ($1$~keV$-10$~MeV), $\mathrm{\mathrm{T_{90}}}$ is used as a proxy for duration and d$_{\rm L}$ is the luminosity distance of the source.\\
(iii) The minimum variability timescale $t_{\rm v}$, that is directly connected to the size of the emitting radius R$_{\rm is}$ through the Eq.~\eqref{eq:is} \citep{guetta}.\\
(iv) The peak value of the gamma-ray energy spectrum E$_{\mathrm{peak}}$.

\subsection{Uncertainties in neutrino flux computation}
\label{uncertainties}
Unfortunately, the intrinsic parameters of the emission 
regions, like the boost Lorentz factor $\Gamma$ and the 
variability timescale $t_{\rm v}$, cannot reliably be 
determined on a source-by-source basis. In few cases the 
Lorentz factor can be estimated: in the so-called \textquoteleft afterglow 
onset method' \citep{sari_L_gamma}, one can relate the energy 
break observed in the GRB light curve during the afterglow 
phase to the jet deceleration time and hence to the initial 
jet speed. Alternatively, one can use the maximum energy of 
observed photons 
\citep{pair_opacity,pair_opacity2,pair_opacity3,pair_opacity4,pair_opacity5} or the quiescent periods between the prompt 
emission pulses, in which the signal of external
shock is expected below the instrument threshold 
\citep{method3}, to infer an average $\Gamma$ of the jet. The former approach was for instance adopted in \citet{lu} for a 
sample of 38 GRBs, from which the authors could derive the 
following correlation between the Lorentz factor $\Gamma$ and the mean isotropic gamma-ray luminosity 
L$_{\gamma,\mathrm{iso}}$:
\begin{equation}
\label{eq:correlation_lu}
\Gamma\simeq 249 (\mathrm{L}_{\gamma,\mathrm{iso},52})^{0.30},
\end{equation}
where L$_{\gamma,\mathrm{iso},52} \equiv 
\mathrm{L}_{\gamma,\mathrm{iso}}/(10^{52}$~erg/s). Therefore, 
by knowing the isotropic luminosity of the burst, it is 
possible to infer the jet Lorentz factor. However, the application of this method is not free from uncertainties, as the isotropic luminosity is also often unknown. In fact, in order to derive L$_{\gamma, \mathrm{iso}}$, the knowledge of the redshift is required (because of the luminosity distance $\rm d_{\rm L}= 
\mathrm{d}_{\rm L}(z)$). As redshift is only known in 11 \% of the cases, a method accounting for the observed redshift distribution of long GRBs was applied in order to estimate respectively i) luminosity distance, ii) isotropic gamma-ray luminosity and iii) bulk Lorentz factor, for each GRB in the selected sample.
Specifically, 1000 random extractions of the $z$ value are performed for GRBs with unknown $z$, according to the redshift distribution of 
long GRBs, as observed by Swift since 2005 and shown in 
Fig.~\ref{fig:swift_distribution}. It is worth mentioning that the introduction of such a distribution in our analysis does not introduce any bias, as it can be shown that the Swift $z$-distribution is representative of the entire sample of long GRBs detected by any instrument from 1997 until today. Nevertheless, the Swift distribution appears very suitable for our purpose, as it can be easily accessed though the satellite's online catalog.
Therefore, for each GRB whose redshift measurement is missing, a value of $z$ is assigned, which allows to first compute the luminosity distance $\rm d_{\mathrm{L}}$ and then L$_{\gamma,\mathrm{iso}}$. Note that the
resulting value of isotropic luminosity is also required 
to be between $10^{49}$ and $10^{54}$~erg/s
since this is the luminosity interval where long GRBs are
detected. Further details on this method and the resulting $\Gamma$ distribution obtained for the selected GRB sample are provided in Appendix~\ref{appendixB}.\\\\
A similar procedure of random extraction according to a known 
distribution of values is adopted for the minimum variability
timescale $t_{\mathrm{v}}$, that is known only in the 33 per 
cent of the cases. For this reason, a distribution of known 
values of $t_{\mathrm{v}}$ for long GRBs, as obtained from 
Fourier analyses on burst light curves \citep{tv1,tv2,tv3}, is 
built as shown in Fig.~\ref{fig:tv_distribution}. For each GRB
with unknown $t_{\mathrm{v}}$, 1000 values of such parameter 
are randomly extracted from this distribution. Note that the 
default value previously adopted in ANTARES GRB search 
\citep{antares_grb1} and advocated in \citet{guetta}, $t_{\mathrm{v}}=10$~ms, is actually 
located in the tail of the measured distribution, that on the
other hand peaks around $0.5$ $\mathrm{s}$. Clearly, the 
default values assumed so far are not representative of the 
different properties of the GRB population.\\\\
Hence, by using the extracted values of redshift $z$ and 
variability timescale $t_{\mathrm{v}}$, 1000 fluxes for each GRB 
(for which $z$ and/or $t_{\mathrm{v}}$ are unknown) are 
simulated, in order to estimate the final neutrino fluence by assuming values of the unknown parameters spanning their allowed ranges. The method allows also to investigate how these uncertainties affect the neutrino spectra and to identify the parameter that contributes the most. Therefore, the following procedure is adopted for those 
sources lacking both $z$ and $t_{\mathrm{v}}$:\\
(i) Calculate the average neutrino fluence resulting from the 1000 simulations.\\
(ii) Use the standard deviation $\sigma$ of the obtained distribution as uncertainty on the average fluence.\\
(iii) Provide the results in terms of $\mathrm{E^2_{\nu_{\mu}}}\mathrm{F_{\nu_{\mu}}} \pm 2\sigma$.\\
When both $z$ and $t_{\mathrm{v}}$ are known (30 GRBs
in the sample), the statistical error around the flux is 
obtained by propagating the measured parameter uncertainties on
$\mathrm{E^2_{\nu_{\mu}}}\mathrm{F_{\nu_{\mu}}}$. In such cases, the
uncertainties are so small that the relative difference between 
$\mathrm{E^2_{\nu_{\mu}}}\mathrm{F_{\nu_{\mu}}}$ and 
$\mathrm{E^2_{\nu_{\mu}}}\mathrm{F_{\nu_{\mu}}} \pm 2\sigma$ 
is negligible, of the order of $10^{-1}$ in the worst cases. However, in few cases, the uncertainty on redshift is not available from measurements: in these cases, the uncertainty has been considered on the last significant digit. In Appendix~\ref{appendixC}, few examples referring to the different cases here explained are reported.\\\\With respect to the correlation adopted in Eq.~\eqref{eq:correlation_lu}, it is worth noting that several expressions of it exist in the literature, which mainly differ in the observational strategy and physical description of the GRB evolution they rely upon. For instance, \citet{ghirlanda_corr} found a relation between $\Gamma$ and the peak luminosity $\rm L_{\gamma,peak}$, by relying on the backwards extrapolation of the self-similar deceleration solution for the shock evolution, as derived by \citet{blandford} (BM). With respect to the method here adopted, the \citet{ghirlanda_corr} approach comes with two further assumptions: i) that in correspondence of the deceleration stage the system dynamics has entered the BM self-similar solution and ii) that the intersection of the two asymptotic power-law phases adopted to describe the shock evolution corresponds to the observed peak time of the afterglow light curve. Because of these stringent limitations, this analysis will adopt the standard approach for the $\Gamma$ estimation by \citetalias{lu}. Clearly, this choice impacts the neutrino flux expectations, in that a significantly different evaluation of the bulk Lorentz factor might lead to a variation in the expected location of the internal shock radius (see Eq.~\eqref{eq:is}). As the neutrino flux is expected to be extremely sensitive to the Lorentz factor \citep{he2012}, a treatment of the additional systematics associated with adopting a different method for deriving $\Gamma$ is presented in Appendix \ref{appendixA}. \\\\

\begin{figure}
\centering
	\includegraphics[width=0.7\columnwidth]{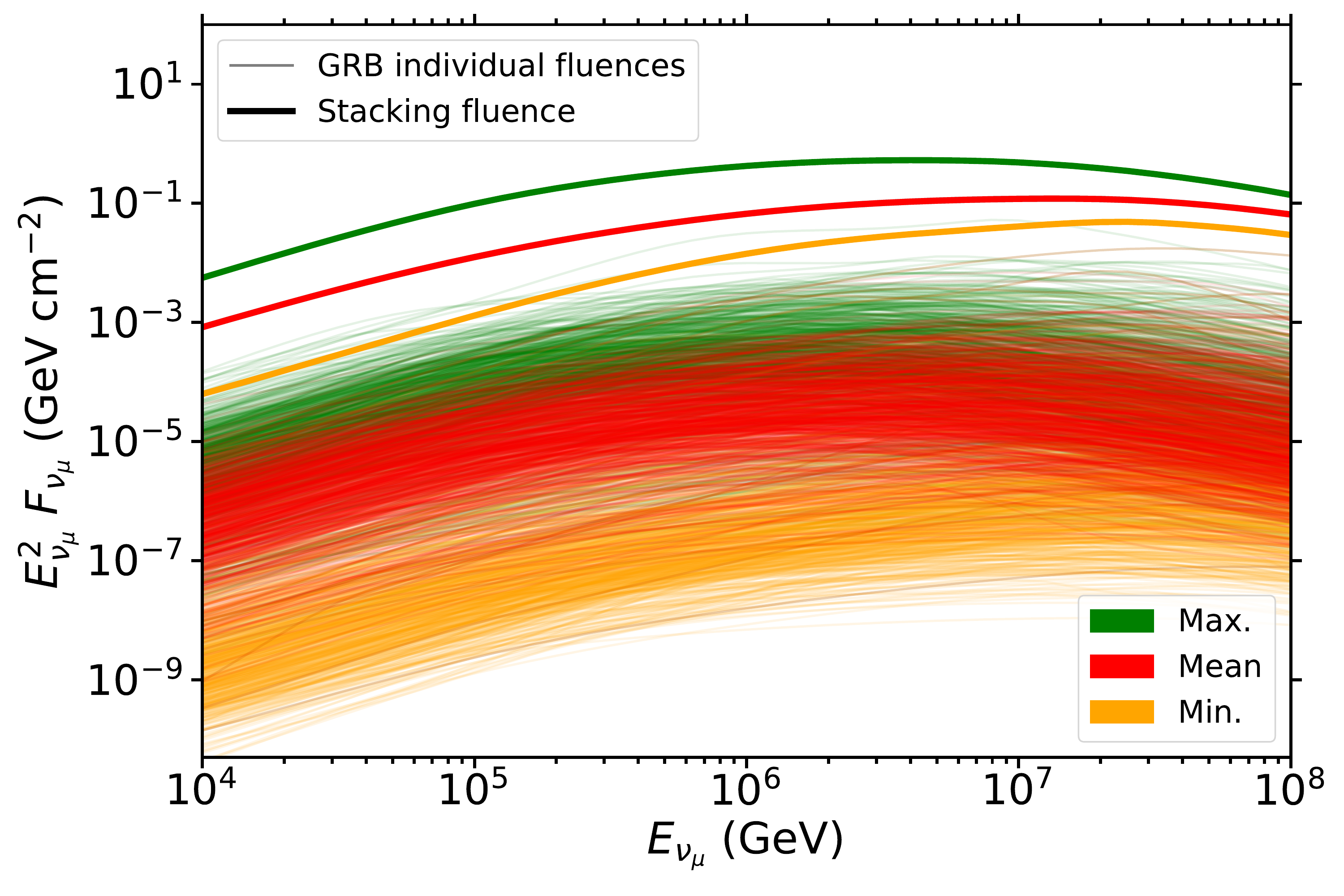}
    \caption{Individual fluences  calculated for each GRB of 
    the 784 in the sample (thin lines) and the corresponding
    stacked fluence (thick line), calculated as in Eq. 
    ~\eqref{stacking}. The mean 
    ($\mathrm{E^2_{\nu_{\mu}}}\mathrm{F_{\nu_{\mu}}}$), 
    mininum ($\mathrm{E^2_{\nu_{\mu}}}\mathrm{F_{\nu_{\mu}}} -
    2\sigma$) and maximum 
    ($\mathrm{E^2_{\nu_{\mu}}}\mathrm{F_{\nu_{\mu}}} + 
    2\sigma$) fluences are shown in red, orange and green, 
    respectively.}
    \label{fig:stacking_all}
\end{figure}

\begin{figure}
\centering
	\includegraphics[width=0.7\columnwidth]{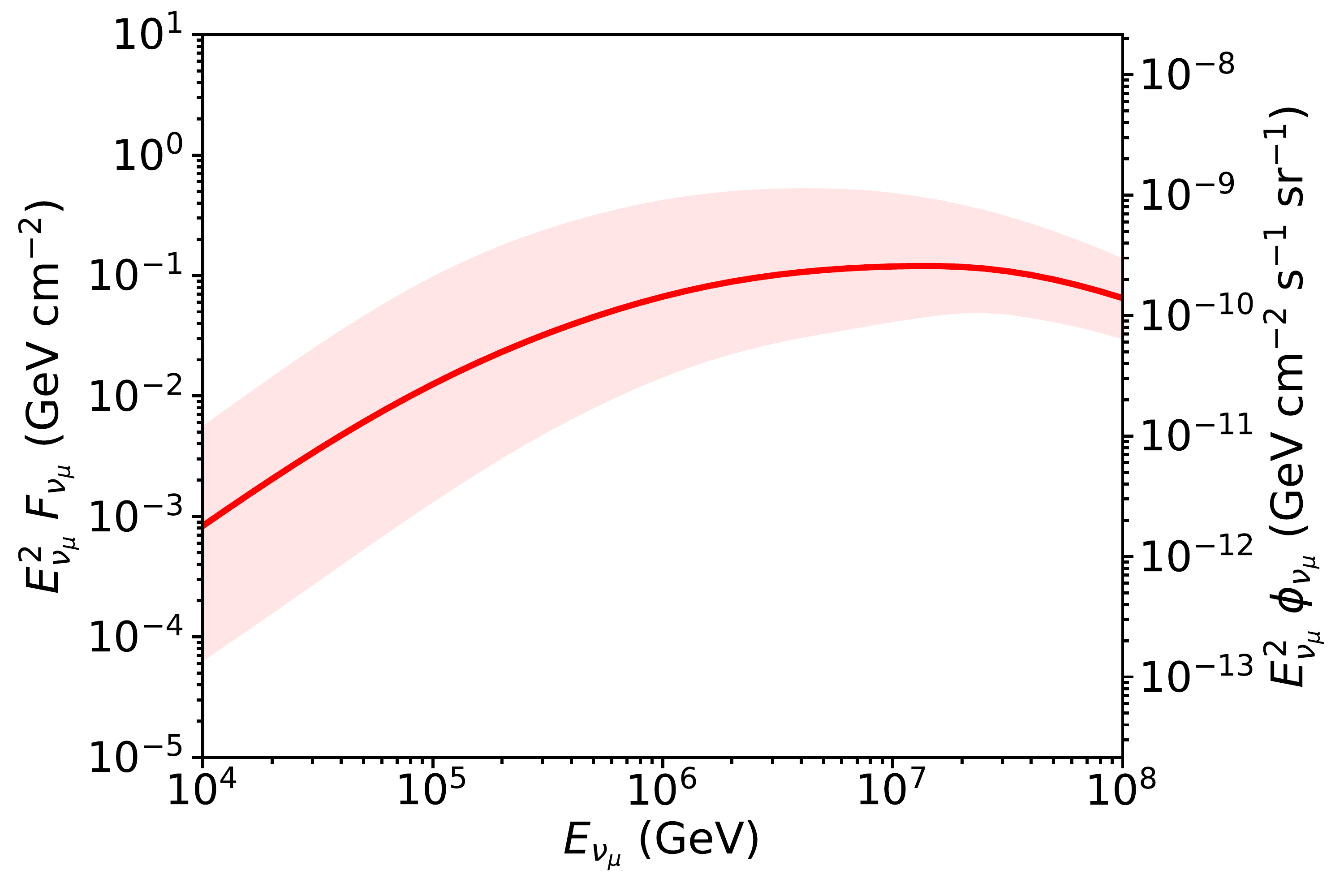}
    \caption{Total neutrino fluence 
    $\mathrm{E^2_{\nu_{\mu}}}\mathrm{F_{\nu_{\mu}}}$ expected 
    from the 784 GRBs in the sample selected in the period 
    2007-2017 (left-hand axis), as in Eq.~\eqref{stacking},
    and corresponding quasi-diffuse neutrino flux 
    $\mathrm{E^2_{\nu_{\mu}}}\mathrm{\phi_{\nu_{\mu}}}$ 
    (right-hand axis), as defined in Eq.~\eqref{quasi-diffuse}. The shaded region indicates the
    error band, obtained from the sum of the individual 
    maximum and minimum fluences for each GRB in the sample 
    (see Fig. \ref{fig:stacking_all}).}
    \label{fig:stacking2}
\end{figure}

\subsection{Cumulative neutrino fluence from all GRBs in the 
sample}
\label{cumulative}
By summing over all the individual neutrino fluences, the 
total fluence expected from the cumulative contribution of the
selected 784 GRBs in the period 2007-2017 is calculated as:
\begin{equation}
\label{stacking}
\mathrm{E^2_{\nu_{\mu}}}\mathrm{\mathrm{F}_{\nu_{\mu}}}=\sum_{
i=1}^{\mathrm{N_{GRB}}=784}(\mathrm{E^2_{\nu_{\mu}}}\mathrm{F_{
\nu_{\mu}}})^{i}.
\end{equation}
In Fig.~\ref{fig:stacking_all}, the expected minimum, mean and
maximum fluences respectively defined as 
$\mathrm{E^2_{\nu_{\mu}}}\mathrm{F_{\nu_{\mu}}}-2\sigma$, 
$\mathrm{E^2_{\nu_{\mu}}}\mathrm{F_{\nu_{\mu}}}$ and 
$\mathrm{E^2_{\nu_{\mu}}}\mathrm{F_{\nu_{\mu}}}+2\sigma$ are shown 
for each GRB and for the whole sample. Focusing on the 
total fluence, note that the maximum and minimum fluences 
define the error band around the mean one, shown in 
Fig.~\ref{fig:stacking2}. It is possible also to convert the 
total neutrino fluence of the sample of $\mathrm{N_{GRB}}$ 
into the quasi-diffuse neutrino flux induced by the same 
sources, by rescaling the total fluence with the average rate 
of GRBs distributed over the full sky expected per year. Hence the quasi-diffuse neutrino flux 
is obtained as

\begin{equation}
\label{quasi-diffuse}
\mathrm{E^2_{\nu_{\mu}}}\mathrm{\phi_{\nu_{\mu}}}=\sum_{i=1}^{
\mathrm{N_{GRB}}}(E^2_{\nu_{\mu}}\mathrm{ F 
_{\nu_{\mu}})}^i\frac{1}{4\pi}\frac{1}{\mathrm{N_{GRB}}}667\mathrm
{yr}^{-1},
\end{equation}
where an annual rate of long GRBs equal to 667 per year
is considered, in agreement with the previous ANTARES analyses
\citep{antares_grb1, antares_grb2}.
The diffuse neutrino flux computed with this method is 
indicated in the right-hand axis of Fig.~\ref{fig:stacking2}. 
This quantity is actually more interesting than the total 
expected fluence, since it allows to compare the neutrino flux
produced by the GRBs in the analysis with both the sensitivity
of neutrino telescopes and the measurement of the 
astrophysical neutrino flux reported by IceCube, in order to 
constrain the contribution of GRBs to this flux (refer to Sec.
\ref{sec:results} for more details).

\section{Signal simulation: the detector probability density function}
\label{sec:detector}
For each source in the sample, a MC simulation of the
expected signal is performed in the so-called run-by-run mode, i.e.
accounting for the specific detector condition at the time that the
GRB occurred, in the same way as in \citet{antares_grb1}. In this 
way, the event generation is able to accurately describe the data 
taking and calibration conditions of the detector during the run in
which each GRB happened. 
Both tracks, resulting from $\nu_{\mu}$ charged current interactions (CC), and showers, produced at $\nu_{\mu}$ neutral current (NC) as well as at $\nu_{e}$ both 
CC and NC interactions, are included in the simulation and signal events are generated from the specific location of the 
sky where the GRB was observed by gamma-ray satellites. To take the
ANTARES absolute pointing uncertainty into account, the GRB local 
coordinates used in the MC signal production are shifted of a quantity randomly generated following \citet{positioning, 
giulia_illuminati} (see also \citet{moon_shadow} and \citet{sun_shadow} for other studies on the ANTARES pointing accuracy).\\
Since only 
GRBs below the ANTARES horizon at the trigger time are considered 
in this search to reduce the atmospheric muon background, 
neutrinos are simulated from the direction of the GRB and passing 
through the Earth, following the simulation scheme described in 
\citet{acceptance}. upward going muon tracks are then reconstructed, to
compute the acceptance of the detector,  with the same algorithm as
in \citet{antares_grb1}. The quality of the reconstruction is 
estimated through two parameters: $\Lambda$, the track-fit quality 
parameter, and $\beta$, the estimated angular uncertainty on the 
muon track direction \citep{quality_cuts}. To improve the 
signal-to-noise ratio, to ensure a good quality 
reconstruction and also to limit the atmospheric muon contamination, only 
tracks with $\beta<1^{\circ}$ are considered in the analysis. The search 
is then optimised through varying a cut on $\Lambda$ selecting tracks above 
a given threshold $\Lambda_{\rm cut}$, as explained in Sec. \ref{sec:optimisation}.
\\The distribution of the 
angular distance between the reconstructed track direction (for
each $\Lambda_{\mathrm{cut}}$) and the GRB's coordinates, normalised to the total number of events, defines the signal 
Probability Density Function (PDF) 
$S(\alpha)=\mathrm{dN}(\alpha)/{\mathrm{d} \Omega}$,
where $\alpha$ is the angular distance between the simulated GRB 
position and the reconstructed muon direction and d$\Omega$ is the 
differential solid angle d$\Omega=2\pi \sin\alpha \rm{d}\alpha$. 
The signal PDF is fitted with a function that is flat for small 
values of $\alpha$ and by a Rayleigh distribution \citep{rayleigh} for larger values.

\section{Background estimation}
\label{sec:background}
\begin{figure}
\centering
	\includegraphics[width=0.7\columnwidth]{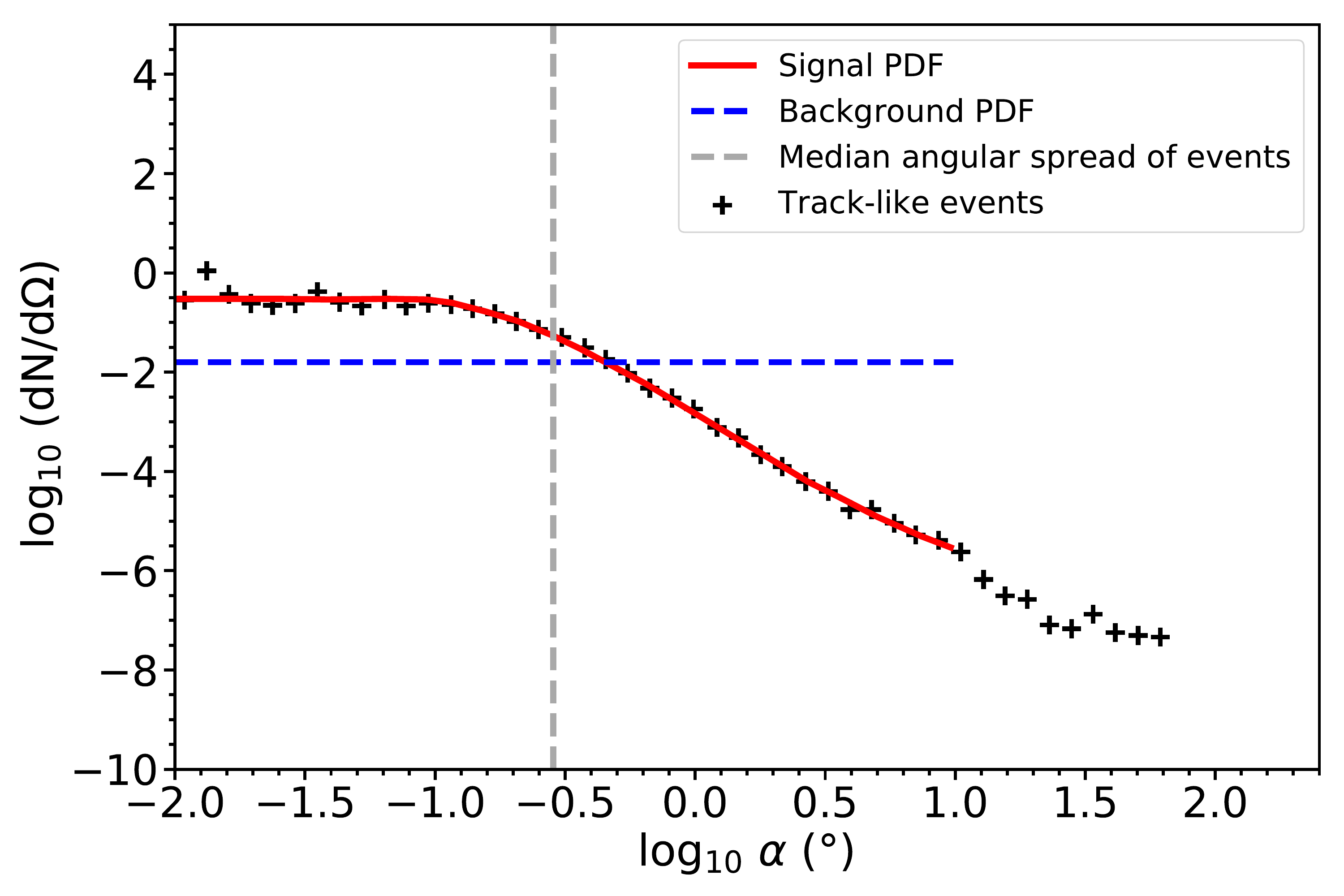}
    \caption{GRB111123A: reconstructed events from the MC signal simulation, per solid angle $\Omega$ as a function of the logarithm of the space angle $\alpha$, obtained with tracks from $\nu_{\mu}$ CC interactions and tracks from $\nu_{\mu}$ NC and $\nu_{\rm e}$ NC+CC interactions (all neutrino channels are shown in black)), with $\beta<1^{\circ}$ and $\Lambda_{\mathrm{cut}}=-5.2$. The vertical dashed line (in gray) indicates the median angular spread of events $\langle \alpha \rangle=0.29^{\circ}$; the horizontal dashed line (in blue) shows the flat background PDF $B(\alpha)$. The red curve is the signal Point Spread Function (PSF), inside the defined angular window, $10^{\circ}$, around the GRB position.}
    \label{fig:fit}
\end{figure}

The  expected number of background events $\mu_{b}$ associated to each GRB, at zenith $\theta$ and azimuth $\phi$, is evaluated directly from data collected by ANTARES off source and off time (between 27th December 2007 and 30th December 2017) as:

\begin{equation}
\label{bkg_rate}
    \mu_b(\theta,\phi)_{\mathrm{GRB}}=1.5~\mathrm{T_{s}}\cdot \langle n(\theta_{\mathrm{GRB}},\phi_{\mathrm{GRB}})\rangle \cdot \cal C,
\end{equation}
where T$_{\rm s}$ is the temporal time window around the GRB occurrence, $\cal C$ is the detector efficiency in the specific runs where each GRB occurred and $\langle n(\theta_{\mathrm{GRB}},\phi_{\mathrm{GRB}})\rangle$ is the time-averaged rate of events reconstructed in the GRB direction.
In the framework of prompt GRB emission, the temporal search window of the neutrino signal was defined in coincidence with the gamma-ray signal, slightly extended to account for uncertainties due to the gamma-ray duration of the event, to the ANTARES data acquisition system and to the propagation time of particles from the satellite to our detector. The time-averaged rate of events reconstructed in the GRB direction, is here estimated with a sample of 15657 runs, equivalent to 61562.5 hours of livetime ($\sim$2565 days). To be conservative, this average value is compared with the mean of time-averaged rates within a $10^{\circ}$ cone around the GRB position, choosing the highest between these two values. This is performed in fact as to account also for the non-uniformity of the background in the vicinity of the GRB position.
Finally, in Eq.~\eqref{bkg_rate} the factor 1.5 is included to conservatively increase the background estimate by $50$ \%.\\
\\The background PDF, $B(\alpha)=\mathrm{d} \rm N(\alpha)/\mathrm{d} \Omega$ is 
assumed to be flat in $\Omega$ within the search cone angle, assuming the 
value as calculated in Eq.~\eqref{bkg_rate}. As a result, the average number of background events expected within a search cone of $10^{\circ}$ 
around a given GRB position is found to be of the order of $10^{-4}$.\\\\For a more detailed description of the signal simulation and background estimation described see \citet{antares_grb1}.\\In Fig.~\ref{fig:fit} the results of the entire analysis chain for a particular GRB (taken as an example), GRB111123A, are presented. The figure shows the signal and background PDFs up to a distance of $10^\circ$ from the simulated GRB position. The signal PDF is obtained by considering all the neutrino events simulated that have been reconstructed as tracks with $\Lambda_{\rm cut}=-5.2$. The median angular spread of events (i.e. the median angular resolution) is also provided.

\section{Maximum likelihood and pseudo-experiments}
\label{sec:pseudo-exp}
MC pseudo-experiments are simulated individually for each GRB with the aim of constructing an ensemble of independent replications of the data acquisition and computing the significance of the measurement. 
\\For each GRB, different sets of simulations are generated by varying $\Lambda_{\mathrm{cut}}$ from $-5$ to $-5.8$. For each of these cuts, $\sim 4 \times 10^6$ signal events and $\sim 4 \times 10^{11}$ background events are simulated. A test statistics $Q$, defined as the ratio between the likelihood in the hypothesis of signal plus background and the likelihood in the background only hypothesis, is evaluated in the form of an 'extended maximum likelihood ratio' \citep{test_statistic}. Furthermore, to determine the statistical significance of measurements, the p-value\footnote{The two-sided convention is used here, namely $p_{3\sigma}=2.7\times10^{-3}$, $p_{4\sigma}=6.3\times10^{-5}$, $p_{5\sigma}=5.7\times10^{-7}$.} is calculated, i.e. the probability to yield $Q$-values at least as high as that observed if the background-only hypothesis was true. At the end of this procedure, the optimal cut on the quality parameter, $\Lambda_{\rm cut}$, is chosen as the one maximising the Model Discovery Potential (MDP), i.e. the probability to observe an excess with a p-value lower than the pre-defined threshold at a given statistical accuracy assuming the signal predicted by the theoretical model (NeuCosmA).\\This strategy was already used by \citet{antares_grb1} and by \citet{antares_bright}. However, there is a difference here in the MDP calculation: the systematic uncertainties in the ANTARES acceptance, that translate into a systematic uncertainty on the value of the estimated signal $\mu_{\rm s}$, are considered in this work, consistently with other previous ANTARES analyses on neutrino sources \citep{acceptance,giulia_illuminati}.

\section{Stacking analysis and search optimisation}
\label{sec:optimisation}
\begin{figure}
    \centering
	\includegraphics[width=0.7\columnwidth]{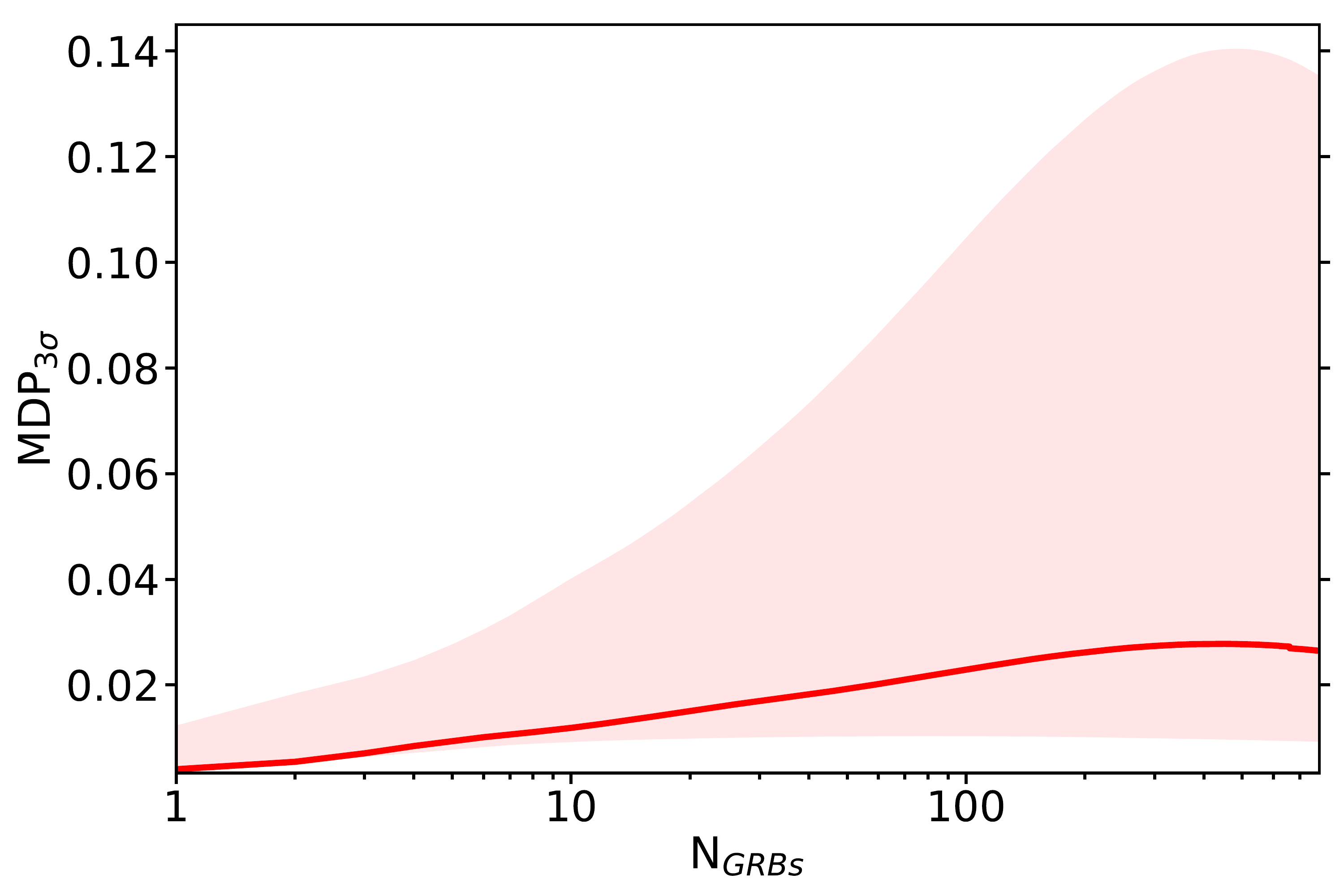}
    \caption{Model Discovery Potential at $3\sigma$, MDP$_{3\sigma}$, as a function of the number of stacked GRBs, $\mathrm{N_{GRB_s}}$. The thick red line indicates the MDP$_{3\sigma}$ obtained with the mean neutrino fluence, while the shaded region is the uncertainty on MDP$_{3\sigma}$ obtained by considering the minimum and maximum fluences (see Fig.~\ref{fig:stacking_all}).}
    \label{fig:mdp}
\end{figure}

The procedure of stacking sources consists into the definition of a GRB sub-sample that includes in the analysis, among the GRBs sample defined in Sec.~\ref{sec:parameter_selection}, as many candidates in terms of neutrino emission as necessary to obtain the best sensitivity. The progressive inclusion of promising GRBs implies the addition not only of the signal but also of the background that they enclose. For this reason the optimal number of sources to stack is found as a compromise between the statistical reduction and the signal gain due to an increasing number of sources in the final sample. In particular, it corresponds to the value which maximises the probability to make a significant discovery (MDP). The procedure, described in details in \citet{antares_grb1}, has been optimised for a 3$\sigma$ significance level. In Fig. \ref{fig:mdp} it is possible to see that the loss in MDP$_{3\sigma}$ is very limited between the use of the whole sample and of an optimal one. Hence the stacking is performed on the whole GRB sample (784 GRBs). Though the search is not optimal in terms of cumulative MDP$_{3\sigma}$, the track quality cut $\Lambda_{\rm cut}$ is set to optimise the MDP$_{3\sigma}$ of individual GRBs. In this regards, the most promising 10 GRBs at $3\sigma$ are reported in Tab.~\ref{tab:opt}, together with the search time window, the optimised cuts and the corresponding expected number of background and signal events.\\The results of the stacking of all 784 sources is presented in Sec.~\ref{sec:results}, corresponding to an 
MDP$_{3\sigma}$ = 0.027 (0.009; 0.136), where the values in parenthesis represent the range of MDP$_{3\sigma}$ values when the model parameters are allowed to vary within 3$\sigma$.

\begin{table}
	\centering
	\caption{Optimisation results obtained with mean fluences: the first ten GRBs with the highest MDP$_{3\sigma}$ are shown, with the corresponding optimised $\Lambda_{\mathrm{cut}}$ value, the expected number of background $\mu_\mathrm{b}$ and signal $\mu_{\mathrm{s}}$ events at $3\sigma$ and the T$_{\mathrm{s}}$. In the last rows, the sum and mean of the values for all 784 GRBs at $3\sigma$ is given. The naming convention of the GRBs is as the same as used by Fermi (see \url{https://heasarc.gsfc.nasa.gov/W3Browse/fermi/fermigbrst.html}).}
	\label{tab:opt}
	\begin{tabular}{lccccc} 
		\hline
		\hline
		GRB & $\Lambda_{\mathrm{cut}}$ & $\mu_{\mathrm{b}}$ & $\mu_{\mathrm{s}}$ & T$_{\mathrm{s}}$ & MDP$_{3\sigma}$\\
		 & & (events) & (events) & (s) & \\
		\hline
		13042732 & $-5.5$ & 5.3$\times10^{-5}$ & 2.2$\times10^{-3}$ & 33.9 & 2.1$\times10^{-3}$\\
		10072809 & $-5.5$ & 9.7$\times10^{-5}$ & 1.1$\times10^{-3}$ & 268.6 & 9.8$\times10^{-4}$\\
		17101079 & $-5.3$ & 1.0$\times10^{-4}$ & 1.0$\times10^{-3}$ & 252.0 & 9.4$\times10^{-4}$\\
		09072071 & $-5.4$ & 1.8$\times10^{-5}$ & 7.8$\times10^{-4}$ & 21.2 & 6.7$\times10^{-4}$\\
		11092889 & $-5.4$ & 4.4$\times10^{-4}$ & 5.1$\times10^{-4}$ & 115.0 & 4.3$\times10^{-4}$\\
		14041606 & $-5.4$ & 5.5$\times10^{-5}$ & 4.2$\times10^{-4}$ & 36.8 & 4.0$\times10^{-4}$\\
		12070780 & $-5.5$ & 7.9$\times10^{-5}$ & 4.1$\times10^{-4}$ & 69.5 & 3.8$\times10^{-4}$\\
		11122865 & $-5.5$ & 4.0$\times10^{-4}$ & 4.4$\times10^{-4}$ & 163.7 & 3.6$\times10^{-4}$\\
		14081078 & $-5.4$ & 7.6$\times10^{-5}$ & 3.7$\times10^{-4}$ & 97.7 & 3.6$\times10^{-4}$\\
		10091081 & $-5.3$ & 5.4$\times10^{-5}$ & 3.4$\times10^{-4}$ & 27.3 & 3.2$\times10^{-4}$\\
		\hline
		all GRBs:\\
		mean & $-5.3$ & 9.4$\times 10^{-5}$ & 3.8$\times 10^{-5}$ & 86.9 & 3.4$\times 10^{-5}$\\
		sum & & 7.3$\times 10^{-2}$ & 3.0$\times 10^{-2}$ & 6.8$\times 10^{4}$ & 2.7$\times 10^{-2}$\\
		\hline
	\end{tabular}
\end{table}

\section{Results and discussion}
\label{sec:results}

\begin{figure}
\centering
	\includegraphics[width=0.7\columnwidth]{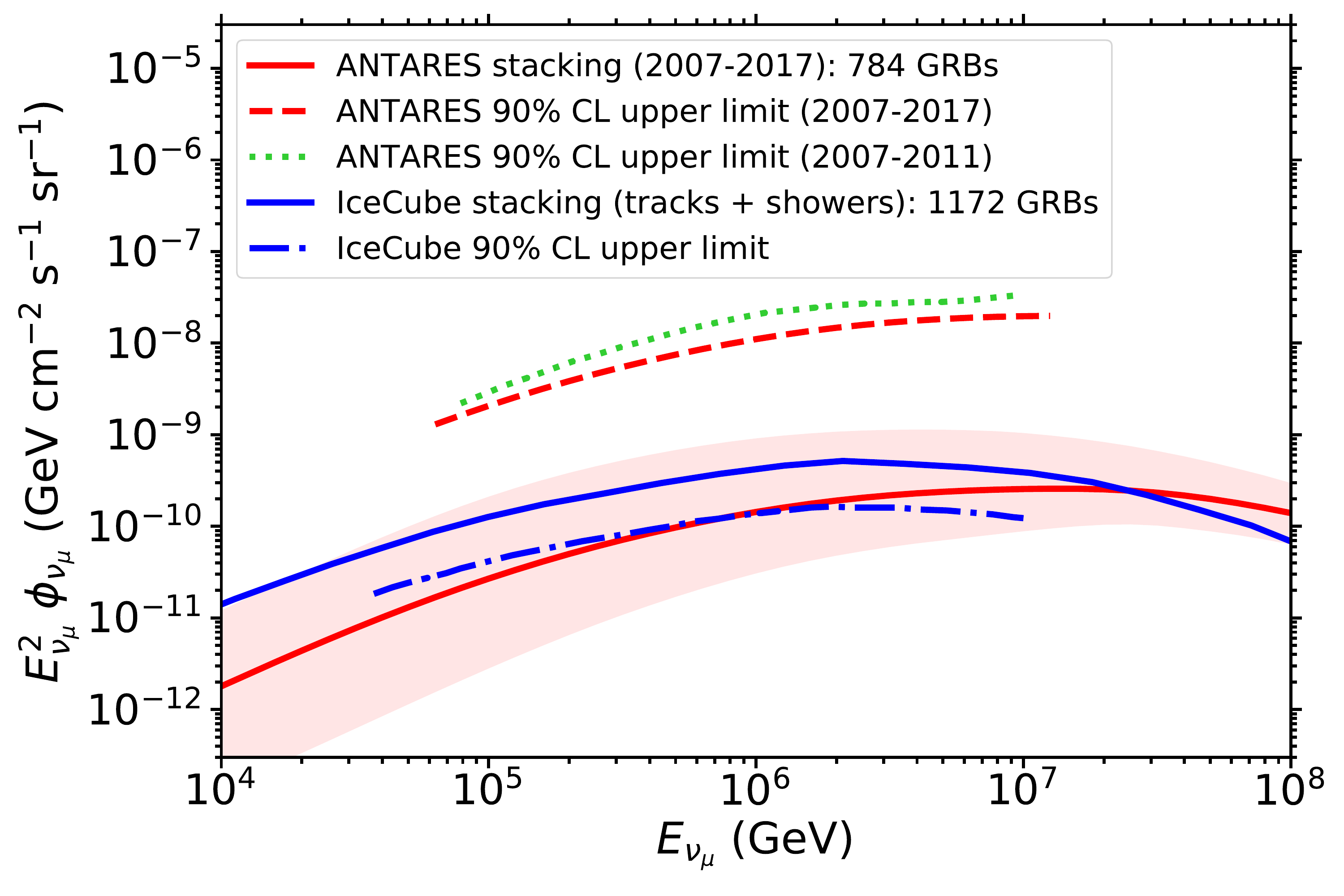}
    \caption{Comparison between the 90 \% CL upper limit (red dashed line) derived by the ANTARES quasi-diffuse flux for 784 GRBs (red solid line), in Eq.~\eqref{quasi-diffuse}, and the previous ANTARES 90 \% CL upper limit (green dashed line) \citep{antares_grb1}. The solid blue line represents the quasi-diffuse flux derived by IceCube for 1172 GRBs and the corresponding dash-dotted blue line shows the corresponding 90 \% CL upper limit \citep{icecube_grb2}.}
    \label{fig:comparison_ul}
\end{figure}

\begin{figure}
\centering
	\includegraphics[width=0.7\columnwidth]{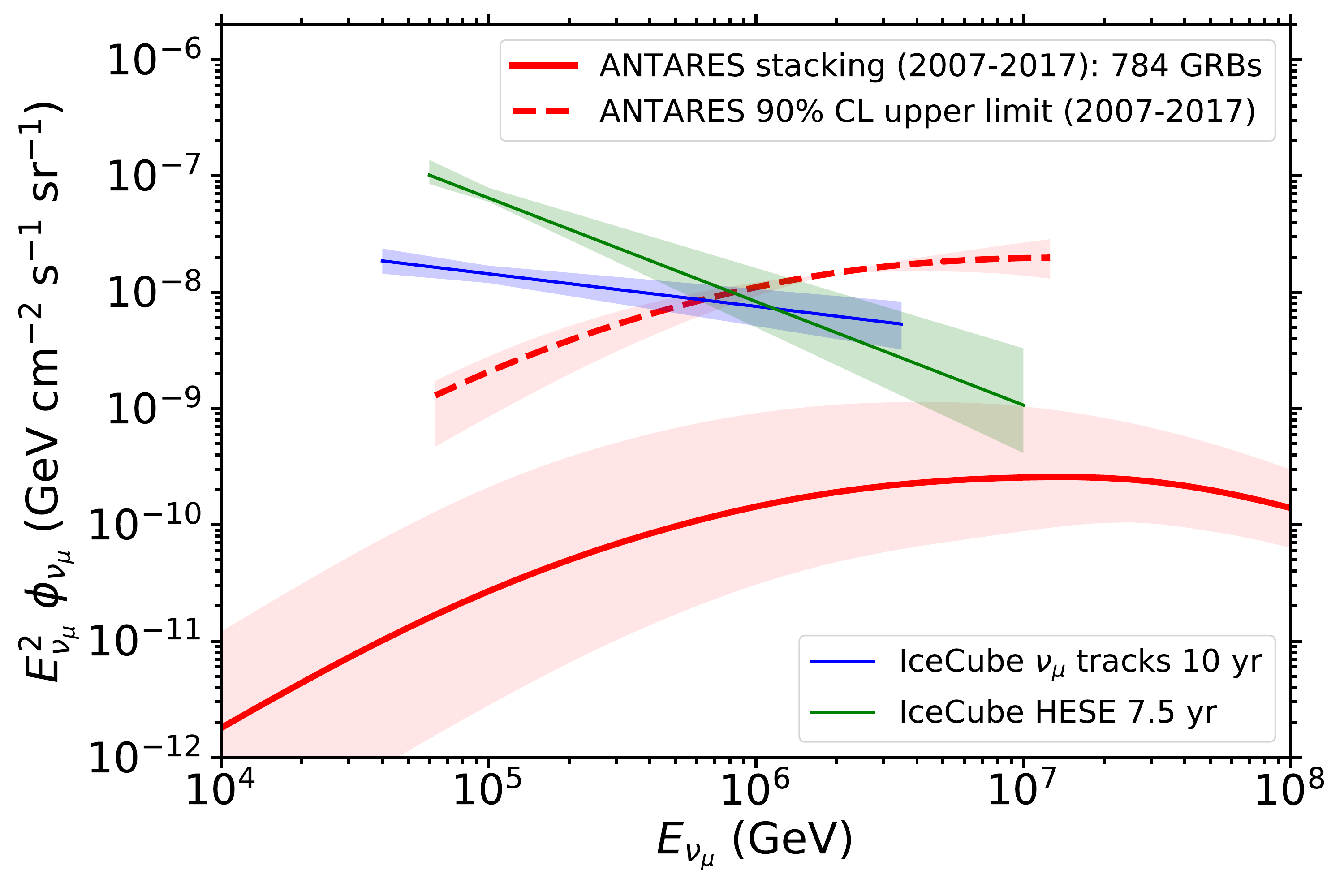}
    \caption{GRB ANTARES quasi-diffuse flux for 784 GRBs, in Eq.~\eqref{quasi-diffuse}, (red solid line) and the corresponding 90 \% CL upper limit (dashed red line). The red shaded regions show the uncertainty around the GRB quasi-diffuse flux, as in Fig.~\ref{fig:stacking2}, and also around the computed upper limit, derived as explained in Sec. \ref{sec:optimisation}. IceCube best fits for $\nu_{\mu}$ tracks in 10 years \citep{icecube_tracks} and for HESE events in 7.5 years of collected data \citep{icecube_hese} are shown in blue and green, respectively.}
    \label{final_plot}
\end{figure}

ANTARES data from the end of 2007 to 2017 are analysed 
according to the cuts identified in the optimisation procedure
presented above, searching for neutrino events in 
spatial and temporal coincidence with the prompt phase of GRBs observed by satellite-based gamma-ray instruments. No neutrino events have passed the selection criteria defined through the optimisation procedure and, thus, no neutrino events are found in spatial and temporal coincidence with the GRB sample, for an equivalent livetime of the search of 18.9~hours. The corresponding 90 \% confidence level (CL) upper limit on the computed neutrino signal $\mathrm{\phi_{\nu_{\mu}}}$ is calculated as

\begin{equation}
\label{ul}
    \mathrm{\phi_{\nu_{\mu}}}^{90\%}=\mathrm{\phi_{\nu_{\mu}}}\frac{\mu_{s}^{90\%}}{\mathrm{n}_\mathrm{s}}=\mathrm{\phi_{\nu_{\mu}}}\frac{2.3}{\mathrm{n}_\mathrm{s}}=\mathrm{\phi_{\nu_{\mu}}} \cdot
    77^{+226}_{-64},
\end{equation}
where the expected number of signal events from the total sample, $\rm n_s$, is estimated to be 

\begin{equation}
\label{eq:number}
\mathrm{n}_\mathrm{s}(\mathrm{N_{GRB}}=784)=0.03^{+0.14}_{-0.02}.
\end{equation}
The factor 2.3 is the 90 \% CL upper limit of the mean of a Poisson process and the value in Eq.~\eqref{eq:number} is a result of the optimisation procedure applied on minimum, mean and maximum fluences, as explained in Sec.~\ref{sec:optimisation}. Note that the relative uncertainty on the expected number of signal events is smaller than the one estimated on the MDP; in other words, the neutrino flux uncertainty due to unknown model parameters is quite limited in the energy range that is relevant for our search. Still, the uncertainty here presented is only partial, as it does not account for the systematics associated with having fixed the correlation in Eq.~\eqref{eq:correlation_lu} to derive the bulk Lorentz factor, which is the parameter expected to most affect the neutrino flux \citep{he2012}. In Appendix \ref{appendixA} such a contribution is also evaluated: as a result of adopting the correlation from \citet{ghirlanda_corr}, the expected neutrino flux is observed shifted to lower energies and with a larger normalization, leading to a significantly larger number of expected neutrino events. However, the experimental cuts obtained with an independent optimization procedure are found to remain almost unaltered. As a consequence, the absence of neutrinos associated to GRBs in ANTARES data allows constraints on both models to be derived. The 90 \% CL upper limits so obtained lay at a comparable level. For the cumulative fluence of Eq.~\eqref{stacking}, this limit reads as $1.3^{+4.1}_{-0.8}\times10^{-2} < \mathrm{E^2_{\nu_{\mu}}}\mathrm{F_{\nu_{\mu}}} < 0.8^{+5.2}_{-0.7}\times10^{-1}$ GeV cm$^{-2}$, in the energy range extending from $6.3\times10^{4}$ GeV to $1.3\times10^{7}$ GeV, which is the region where 90 \% of the mean fluence is expected to be detected by ANTARES. The fluence limit translates into $1.3^{+0.4}_{-0.8}\times10^{-9} < \mathrm{E^2_{\nu_{\mu}}}\mathrm{\phi_{\nu_{\mu}}} < 1.0^{+0.9}_{-0.5}\times10^{-8}$ GeV cm$^{-2}$ s$^{-1}$ sr$^{-1}$ in terms of quasi-diffuse flux (cfr Eq.~\eqref{quasi-diffuse}). The quasi-diffuse expected flux and corresponding upper limit, as calculated from the mean expected fluence, are shown in Fig.~\ref{fig:comparison_ul} and compared to previous ANTARES limits \citep{antares_grb1}. An improvement by a factor $\sim$ 2 on the 90 \% CL upper limit can be observed, due to the increased sample statistics, jointly with having here adopted a more realistic model for neutrino predictions including a detailed study on the model parameters. The results are also compared with the latest IceCube all-sky search \citep{icecube_grb2}, where no statistically significant signal was found by combining both track and shower events for 1172 GRBs. From this comparison, it is possible to appreciate thatthe GRB-neutrino flux expected by IceCube is consistent with the one presented in this work over the entire energy range 10$^4$-10$^8$~GeV, the former being on average higher than the latter due to the larger sample size. The same spectral trend is reflected in individual upper limits. 
It is worth keeping in mind that when comparing results from different analyses, one should consider that the spectral and limit shapes depend on the selected sample, the measured parameters of each burst and their uncertainty, namely the set of parameters that are introduced in the chosen model. Here, for the first time, no default value for the model parameters are used and more physical and realistic values are considered (see Sec~\ref{uncertainties}).\\
Finally, the expected quasi-diffuse neutrino flux from the selected 784 GRBs and the corresponding upper limit can be compared with the diffuse astrophysical flux observed by IceCube. To this extent, Fig.~\ref{final_plot}, provides the IceCube best fits of the neutrino flux, in both the 10~years $\nu_{\mu}$ track data sample \citep{icecube_tracks}, and the 7.5~years High-Energy Starting Events (HESE)\footnote{The neutrino interaction vertex is located inside the detector and its energy is larger than 20~TeV.} sample \citep{icecube_hese}. To allow a more significant comparison, the upper limit derived from this search is reported with its error band (see Eq.~\eqref{ul}). By comparing the ANTARES upper limit with the diffuse astrophysical neutrino flux observed by IceCube, it is possible to conclude that GRBs are not the main contributors to the observed flux below E$_\nu \sim 1$~PeV, within the NeucosmA model framework set with benchmark baryonic loading ($f_{\rm p}=10$). This result confirms previous searches performed by IceCube \citep{icecube_grb3,icecube_grb1,icecube_grb2}. In particular, in the energy region where ANTARES is most sensitive, i.e. below $\sim 100$~TeV, GRBs do not contribute by more than 10 \%. Consequently, the parameter space still allowed to the internal shock model is characterized by sizeably smaller baryonic loading of GRB jets.\\\\It is worth highlighting that this analysis accounts for the contribution to the observed diffuse astrophysical neutrino flux of long resolved GRBs (i.e. triggered). A potentially interesting contribution is constituted by the many GRBs that elude detection (due to their low photon flux) and which is here left unconstrained. As estimated e.g. by \citet{liu}, the neutrino flux from such unresolved GRBs might even be larger than the one due to resolved ones. In addition to this, other interesting classes of sources possibly contributing to the diffuse astrophysical neutrino flux detected by IceCube are:\\
(i) low-luminous GRBs (LLGRBs) (e.g. \citet{murase_ll,llgrbs2}), namely GRBs characterized by a luminosity $\lesssim 10^{49}$~erg~s$^{-1}$;\\
(ii) choked GRBs, which being opaque to radiation in the GeV–TeV band might show up as neutrino sources hidden with respect to gamma-ray observations (e.g. \citet{choked1,llgrbs_neutrinos,choked2}).


\section{Summary and conclusions}
\label{sec:conclusions}
Using ANTARES data from the end of 2007 to 2017, a search for upward going muon neutrinos and anti-neutrinos in spatial and temporal coincidence with 784 GRBs has been performed. The numerical model NeuCosmA was used to estimate the expected neutrino flux from each burst individually, in the context of one-zone internal shock model. A novel aspect of the search here presented is the inclusion in the data analysis chain of the uncertainty that possible unknown parameters, related to the characteristic activity of the central engine, can introduce in the neutrino flux evaluation. This is crucial in order to correctly interpret the validity of model-dependent results, in terms of upper limits set by non-detections of neutrinos in coincidence with GRBs \citep{antares_grb1,icecube_grb2}. These parameters have been identified in the bulk Lorentz factor, variability timescale and source redshift, all of which are affecting the so-called dissipation radius, where shell collisions are realized. Among these parameters, the former was shown to impact the most GRB-neutrino flux predictions. At the same time, it is also possible to marginalize the uncertainty related to it by assuming a correlation with the source isotropic gamma-ray luminosity (which is in turn a physical observable). This was realized by relying upon the observational correlation found by \citetalias{lu}. As a result of such procedure, the minimum variability timescale was found to contribute more than redshift to the uncertainty on the neutrino flux predictions from GRBs. Indeed, when letting $t_{\rm v}$ free to vary, the estimated uncertainty on the neutrino flux expected from the model is observed to span up to several orders of magnitude. As a consequence, the expected $\nu$-fluxes are provided with an uncertainty band of $\pm 2\sigma$. Analogously to previous ANTARES searches \citep{antares_grb1,antares_bright,celliGRB}, MC simulations of the signal predicted by NeuCosmA were performed, while the respective background was estimated directly from off-source data collected by ANTARES. Only track-like events reconstructed within $10^\circ$ in radius from the expected GRB position were selected, and in temporal correlation with the prompt gamma-ray emission. \\
The analysis was optimised on a burst-by-burst basis so as to maximise the discovery potential of the search, thus allowing the identification of the most promising GRBs for ANTARES. However, because a negligible reduction of the MDP$_{3\sigma}$ would have been obtained when stacking the entire catalog, the flux from the whole sample of 784 GRBs was investigated. After unblinding ANTARES data occurred in space and time correlation with GRBs, no event was found to pass the selection criteria, and limits on the contribution of the detected GRB population to the neutrino quasi-diffuse flux were derived. The limits obtained on the cumulative neutrino fluence E$_{\nu_{\mu}}^2F_{\nu_{\mu}}$, relative to the predictions of NeuCosmA, are 
$1.3^{+4.1}_{-0.8}\times10^{-2}$ GeV cm$^{-2}$ and
$0.8^{+5.2}_{-0.7}\times10^{-1}$ GeV cm$^{-2}$, corresponding to $1.3^{+0.4}_{-0.8}\times10^{-9}$ GeV cm$^{-2}$ s$^{-1}$ sr$^{-1}$ and $1.0^{+0.9}_{-0.5}\times10^{-8}$ GeV cm$^{-2}$ s$^{-1}$ sr$^{-1}$, respectively, in terms of quasi-diffuse flux E$_{\nu_{\mu}}^2\phi_{\nu_{\mu}}$ in the energy range from $\sim 60$~TeV to $\sim 10$~PeV. For the sake of completeness, an upper limit was also calculated relatively to the analysis that assumes the \citet{ghirlanda_corr} correlation as a reference model, and it was found to be at a comparable flux level of the one presented here.\\
With these results, ANTARES data provide a further and independent constrain on the contribution of GRBs to the astrophysical neutrino flux. In particular, within standard assumptions of energy partition among accelerated hadrons, leptons and magnetic fields (baryonic loading equal to 10), GRBs are not the main sources of the astrophysical neutrino flux, possibly contributing for less than 10 \% at energies around 100~TeV.

\section*{Acknowledgements}
The authors acknowledge the financial support of the funding agencies:
% France:
Centre National de la Recherche Scientifique (CNRS), Commissariat \`a
l'\'ener\-gie atomique et aux \'energies alternatives (CEA),
Commission Europ\'eenne (FEDER fund and Marie Curie Program),
Institut Universitaire de France (IUF), LabEx UnivEarthS (ANR-10-LABX-0023 and ANR-18-IDEX-0001),
R\'egion \^Ile-de-France (DIM-ACAV), R\'egion
Alsace (contrat CPER), R\'egion Provence-Alpes-C\^ote d'Azur,
D\'e\-par\-tement du Var and Ville de La
Seyne-sur-Mer, France;
% Germany: 
Bundesministerium f\"ur Bildung und Forschung
(BMBF), Germany; 
% Italy
Istituto Nazionale di Fisica Nucleare (INFN), Italy;
% Netherlands
Nederlandse organisatie voor Wetenschappelijk Onderzoek (NWO), the Netherlands;
% Russia
Council of the President of the Russian Federation for young
scientists and leading scientific schools supporting grants, Russia;
% Romania
Executive Unit for Financing Higher Education, Research, Development and Innovation (UEFISCDI), Romania;
% Spain
Ministerio de Ciencia, Innovaci\'{o}n, Investigaci\'{o}n y Universidades (MCIU): Programa Estatal de Generaci\'{o}n de Conocimiento (refs. PGC2018-096663-B-C41, -A-C42, -B-C43, -B-C44) (MCIU/FEDER), Severo Ochoa Centre of Excellence and MultiDark Consolider (MCIU), Junta de Andaluc\'{i}a (ref. SOMM17/6104/UGR and A-FQM-053-UGR18), 
Generalitat Valenciana: Grisol\'{i}a (ref. GRISOLIA/2018/119), Spain; 
% and GenT (CIDEGENT/2018/034) programs
% Marocco
Ministry of Higher Education, Scientific Research and Professional Training, Morocco.
% A.O.B.:
We also acknowledge the technical support of Ifremer, AIM and Foselev Marine
for the sea operation and the CC-IN2P3 for the computing facilities.

%%%%%%%%%%%%%%%%%%%%%%%%%%%%%%%%%%%%%%%%%%%%%%%%%%

%%%%%%%%%%%%%%%%%%%% REFERENCES %%%%%%%%%%%%%%%%%%

%%%%%%%%%%%%%%%%%%%%%%%%%%%%%%%%%%%%%%%%%%%%%%%%%%

%%%%%%%%%%%%%%%%% APPENDICES %%%%%%%%%%%%%%%%%%%%%
\appendix
\counterwithin{figure}{section}
\counterwithin{equation}{section}

\section{Determining the bulk Lorentz Factor}
\label{appendixB}

\begin{figure}
\subfigure[\label{fig:comparison_julia}]{\includegraphics[width=0.55\columnwidth]{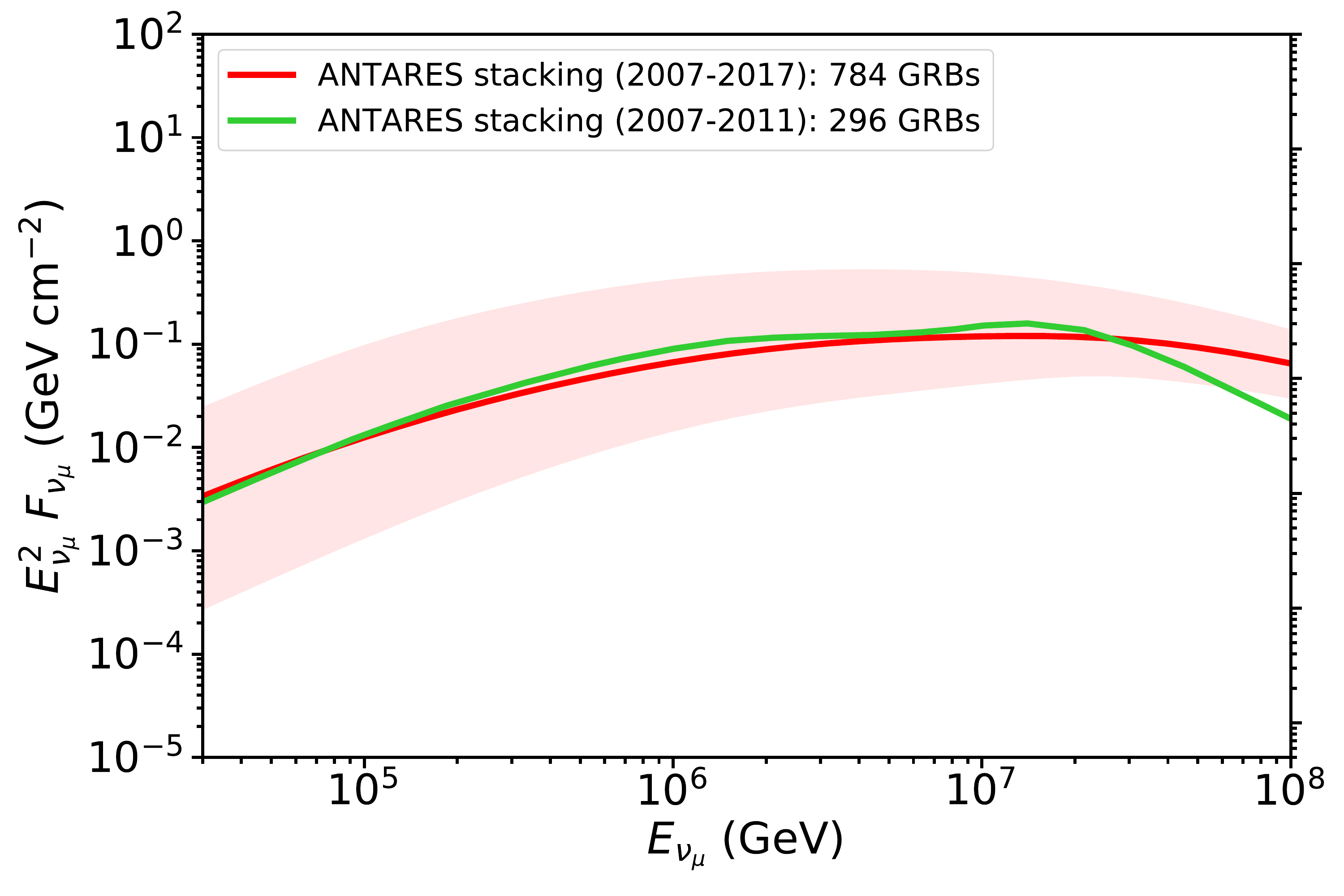}}
\subfigure[\label{fig:gamma_distr}]{\includegraphics[width=0.55\columnwidth]{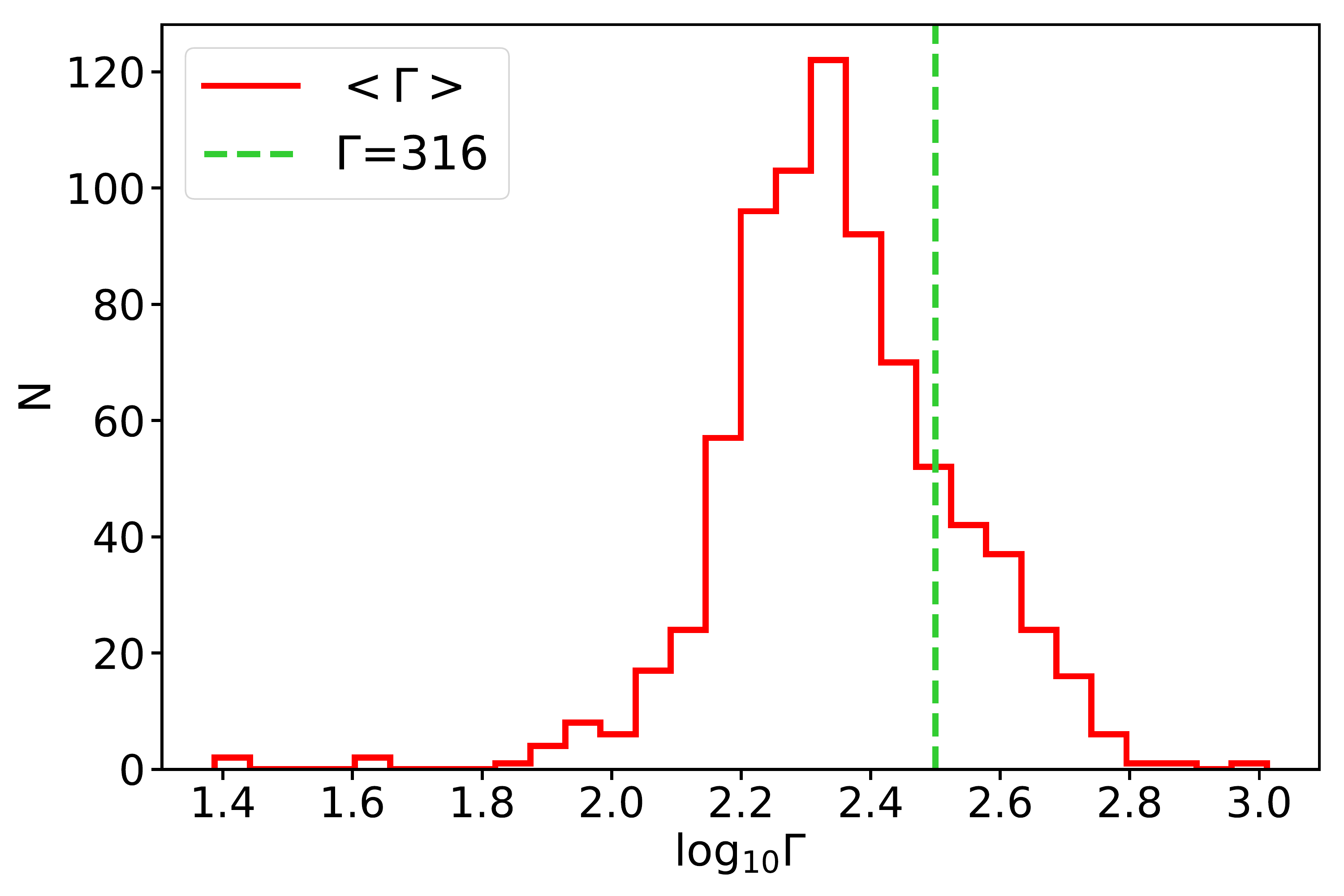}}
\caption{(a) Comparison between the cumulative neutrino fluence expected from the stacking of 784 GRBs in the period 2007-2017 (in red) and the cumulative neutrino fluence obtained in \citet{antares_grb1} from stacking 296 GRBs in the years 2007-2011 (in green) The red shaded region indicates the error band around the neutino fluence estimated in this work, taking into account the several uncertainties affecting the neutrino production in GRBs. (b) Logarithmic distribution of the average bulk Lorentz factor $\langle \Gamma \rangle$ for any burst in the sample (in red), in comparison with default value $\Gamma=316$ (dashed green line) previously used by \citet{antares_grb1}. Note that where a measurement of $z$ was missing $\langle \Gamma \rangle$ is obtained by averaging the 1000 values of redshift, possibly extracted for each GRB. On the other hand, for GRB with measured $z$, a single contribution of $\Gamma$ is present in this plot, as given by Eq.~\eqref{eq:correlation_lu}.}\label{first}
\end{figure}

\begin{figure}
\subfigure[\label{fig:z_distr_spec}]{\includegraphics[width=0.55\columnwidth]{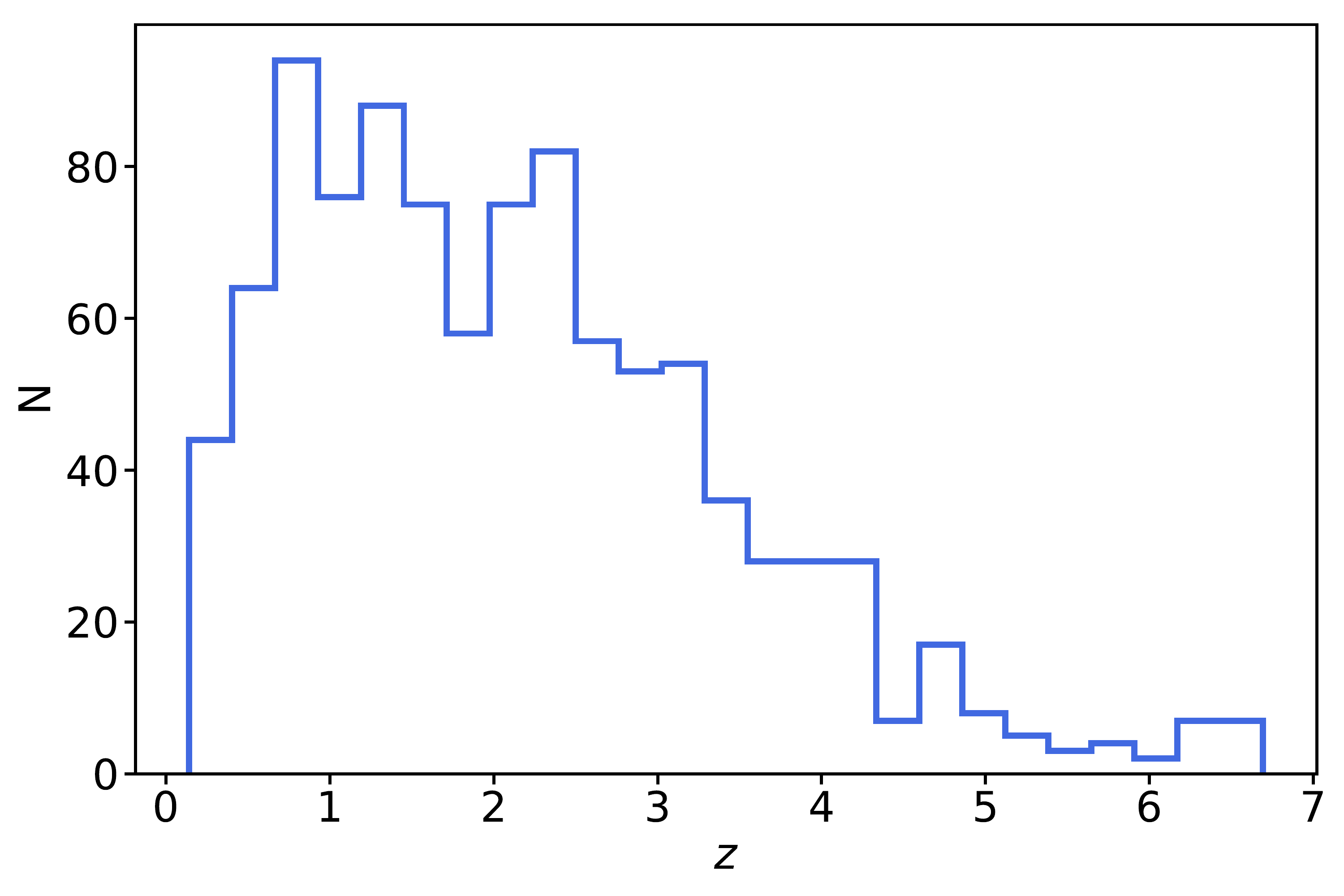}}
\subfigure[\label{fig:gamma_distr_spec}]{\includegraphics[width=0.55\columnwidth]{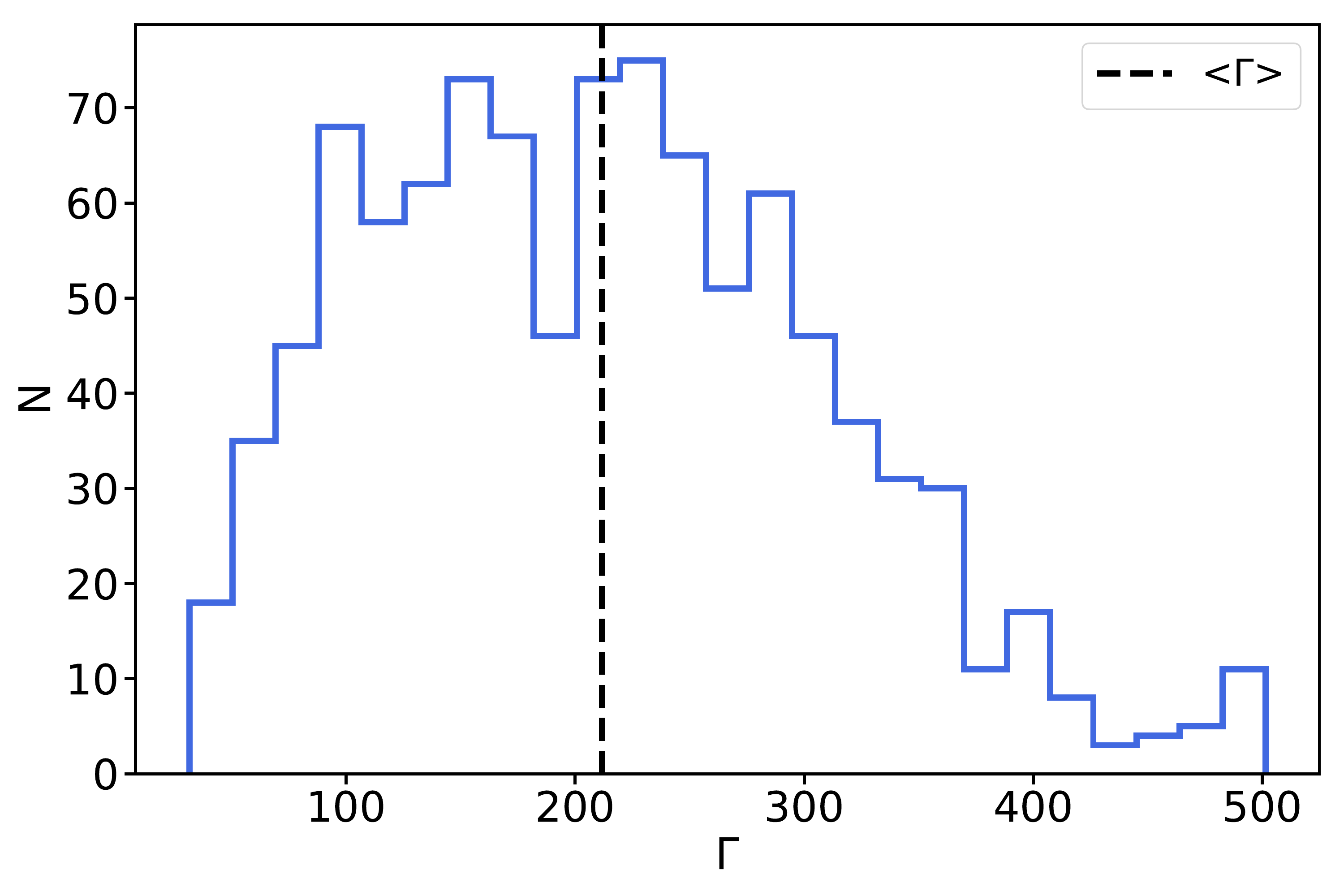}}
    \caption{(a) Distributions of the redshift $z$ values randomly extracted for GRB08102853. (b) Corresponding bulk Lorentz factor $\Gamma$ values obtained by using the correlation in Eq.~\eqref{eq:correlation_lu} \citepalias{lu}. The black dashed line shows the average $\Gamma$ of the considered GRB, $\langle \Gamma \rangle \simeq 210$.}
    \label{first2}
\end{figure}

The bulk Lorentz factor $\Gamma$ of the stellar ejecta is a key parameter to understand the physics of GRBs, extremely powerful sources with an intrinsic mildly relativistic nature. In the standard fireball scenario, the temporal evolution of the jet's speed can be approximated as an initial acceleration phase, followed by a period with $\Gamma$ constant before reaching the external medium and decelerating in it \citep{zhang}.\\
The bulk Lorentz factor determines the frequency of plasma shell collisions, and consequently the rate of particle acceleration. $\Gamma$ affects the shape of neutrino spectra and in particular the spectral breaks.
The first derivations of the energy breaks were performed by \citet{guetta}, who predicted two energy breaks in the neutrino spectra at the energies

\begin{equation}
\label{eq:break1}
\epsilon _{\nu,1}\propto(1+z)^{-2}~\Gamma^2_{2.5}~\epsilon_{\gamma,\rm MeV}^{-1}
\end{equation}and

\begin{equation}
\label{eq:break2}
\epsilon_{\nu,2}\propto(1+z)^{-1}~\Gamma^2_{2.5}~\rm R_{is}~L_{\gamma,52}~\epsilon_{B}^{-1/2},
\end{equation}
where $\Gamma_{2.5}=\Gamma/(10^{2.5})$ and $\epsilon_{\gamma,\rm MeV}=\epsilon_{\gamma}/\rm{MeV}$ is the photon energy. The first break, in Eq.~\eqref{eq:break1}, is due to the synchrotron break observed in the photon spectrum and the second one, in Eq.~\eqref{eq:break2}, comes from the onset of cooling losses in high-energy muons.
Within the model implemented in NeuCosmA (see Sec.~\ref{sec:model}), a third break is expected in the combined $\nu_{\mu}+\bar{\nu}_{\mu}$ spectrum, due to the onset of cooling losses in pions \citepalias{neucosma2}.\\
The stochastic nature of GRBs, in addition with the complex dynamical evolution of the jet, makes it hard to reliably determine a bulk Lorentz factor. In the previous ANTARES search, as well as in several IceCube searches, a default value of $\Gamma=316$ was used \citep{antares_grb1}. In Fig.~\ref{first}\ref{fig:comparison_julia} the stacking fluence obtained in this work is compared with the previous ANTARES estimation, both computed with one-zone modelling of NeuCosmA. However, in this work, a novel method for the estimation of $\Gamma$ is presented, consisting into exploiting the observed correlation among $\Gamma$ and the burst's isotropic luminosity, as found by \citetalias{lu} and reported in Eq.~\eqref{eq:correlation_lu}. Nonetheless, such a correlation cannot be used straightforwardly in most of the cases, since it would require the knowledge of the redshift for each GRB of the sample. Unfortunately redshift is unknown in 90 per cent of the cases: in this situation, for each GRB with $z$ not measured, up to 1000 values of redshift are randomly extracted from a redshift distribution that follows that of long GRBs detected since 2005 by the Swift satellite (see Fig.~\ref{fig:swift_distribution}). Then, from such 1000 values of $z$, 1000 values of bulk Lorentz factor are calculated through Eq.~\eqref{eq:correlation_lu}. By averaging the resulting 1000 values of $\Gamma$ for an individual GRB, $\langle \Gamma \rangle$ is obtained. The resulting cumulative neutrino fluence is shown in Fig.~\ref{first2}\ref{fig:comparison_julia}, where it is also compared with the expected neutrino fluence estimated by the previous ANTARES analysis \citep{antares_grb1}. The two are observed at a comparable level, even though the latest analysis has more than twice more sources than the previous. This result is in fact a consequence of the neutrino modeling adopted: while past predictions tended to overestimate the expected flux by assuming standard values for model parameters, here an accurate modeling is realized by accounting for variations in these parameters reflecting the properties of observed GRBs. An example is given in Fig.~\ref{first2}\ref{fig:gamma_distr}, where the distribution of the $\langle \Gamma \rangle$ values obtained for each burst is shown and compared with the standard value used in the past. The obtained distribution peaks at a value lower than $\Gamma=316$.\\
An example of the procedure just explained is shown for GRB08102853 in Fig.~\ref{first2}\ref{fig:z_distr_spec} and Fig.~\ref{first2}\ref{fig:gamma_distr_spec}, where the redshift and subsequent Lorentz factor distributions are shown, respectively.\\
Moreover, it is worth to recall that in this work measured values of the minimum variability timescale are also used when available, or they are extracted from a distribution of known values. This is another difference with respect to what was assumed in the previous ANTARES search where $t_{\rm v}=10$~ms was considered for all GRBs irrespective of their actual light curve. These differences do have an impact, as shown in this work, on the neutrino spectral shape in comparison with previous analyses.


\section{Individual neutrino fluence simulations}
\label{appendixC}

\begin{figure}
\subfigure[\label{fig:both_unknown}]{\includegraphics[width=0.55\columnwidth]{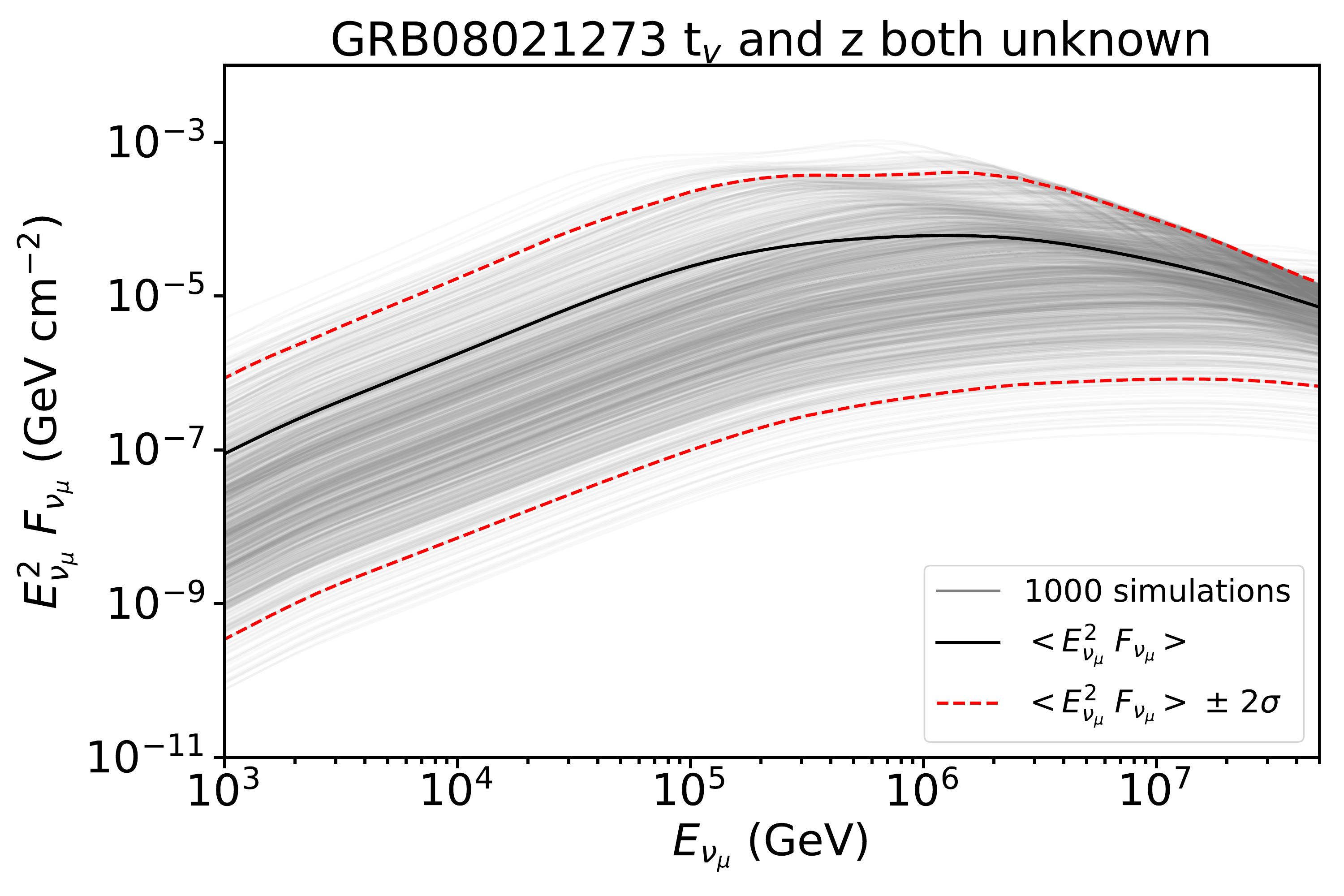}}
\subfigure[\label{fig:tv_unknown}]{\includegraphics[width=0.55\columnwidth]{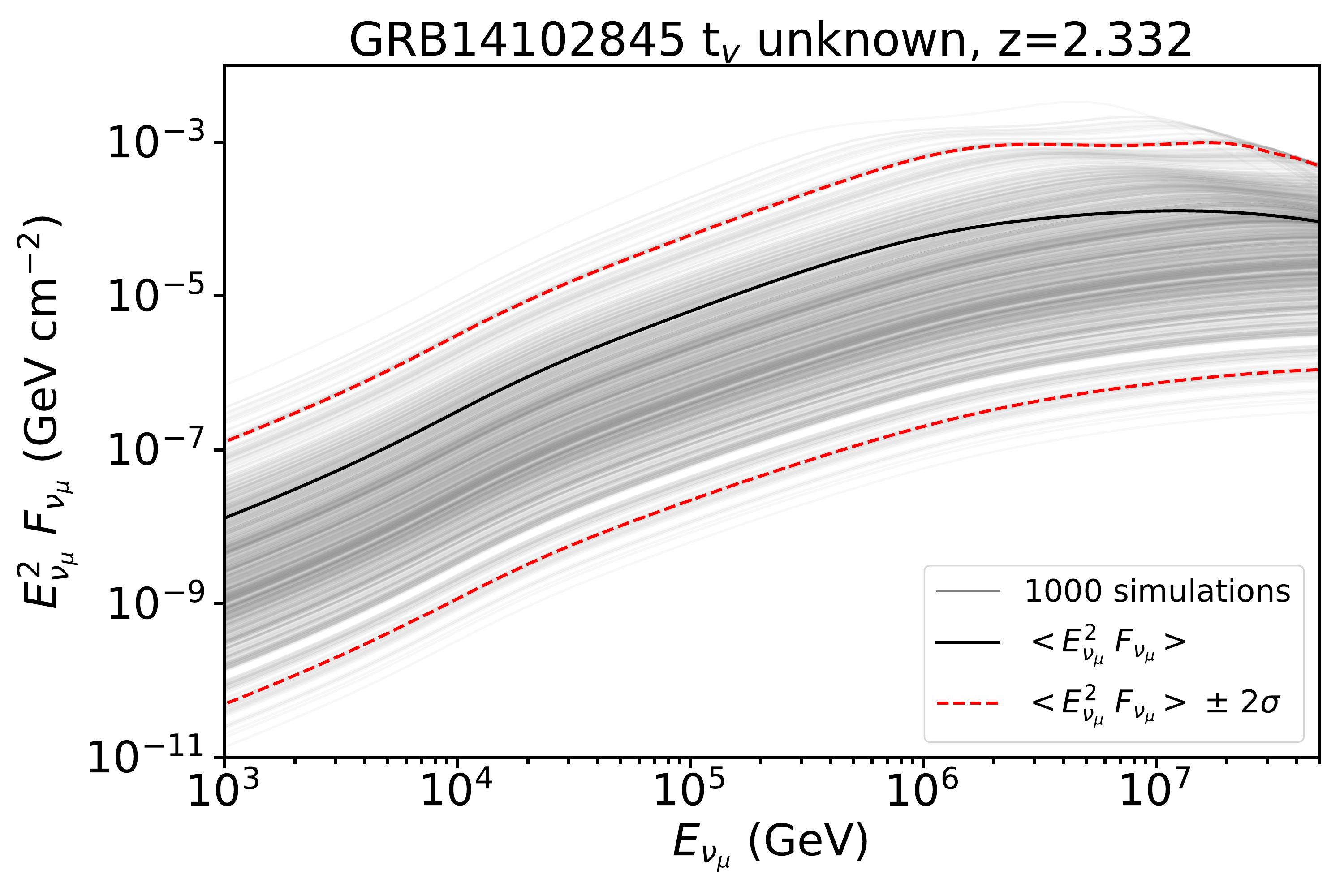}}
\subfigure[\label{fig:z_unknown}]{	\includegraphics[width=0.55\columnwidth]{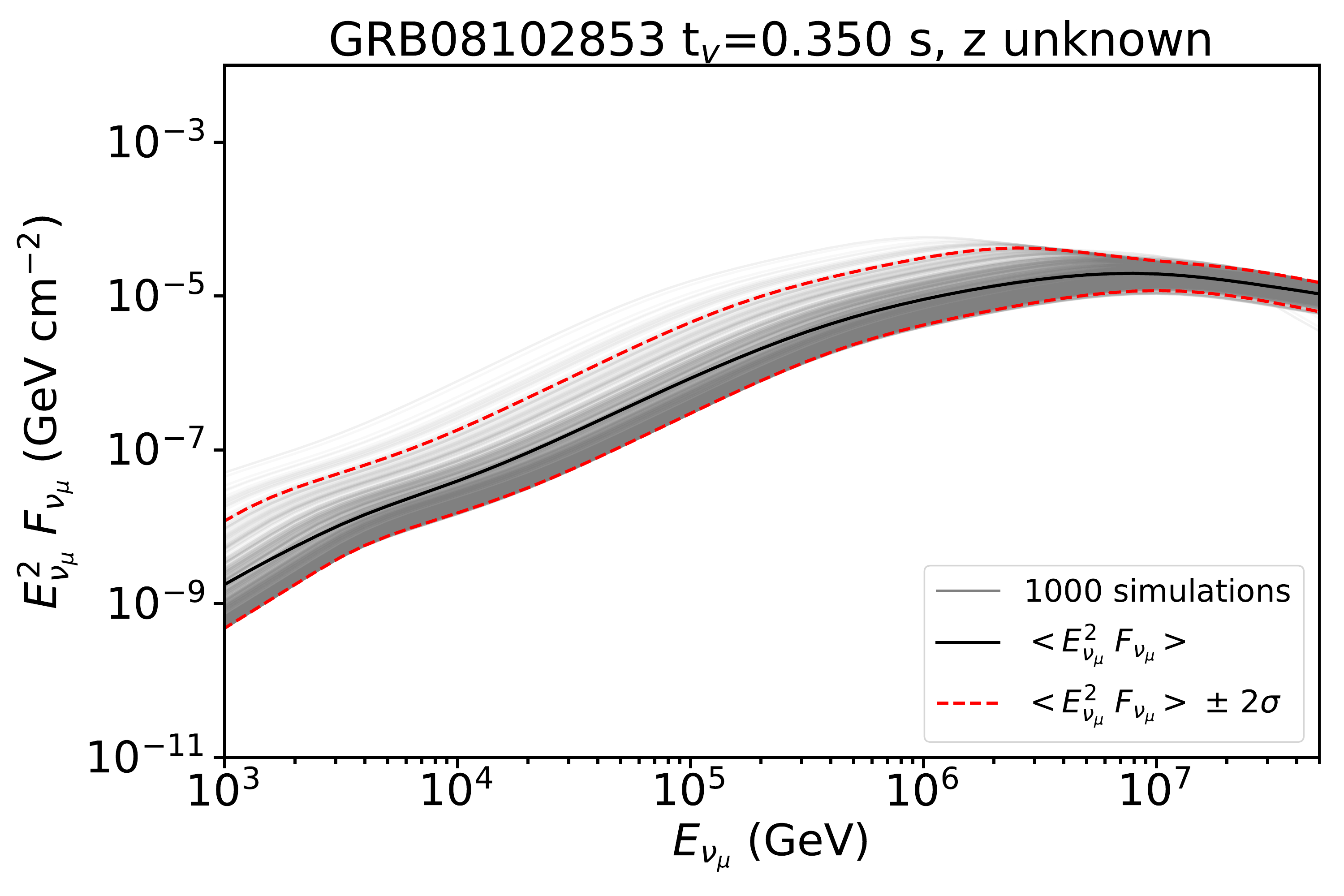}}
\subfigure[\label{fig:both_known}]{\includegraphics[width=0.55\columnwidth]{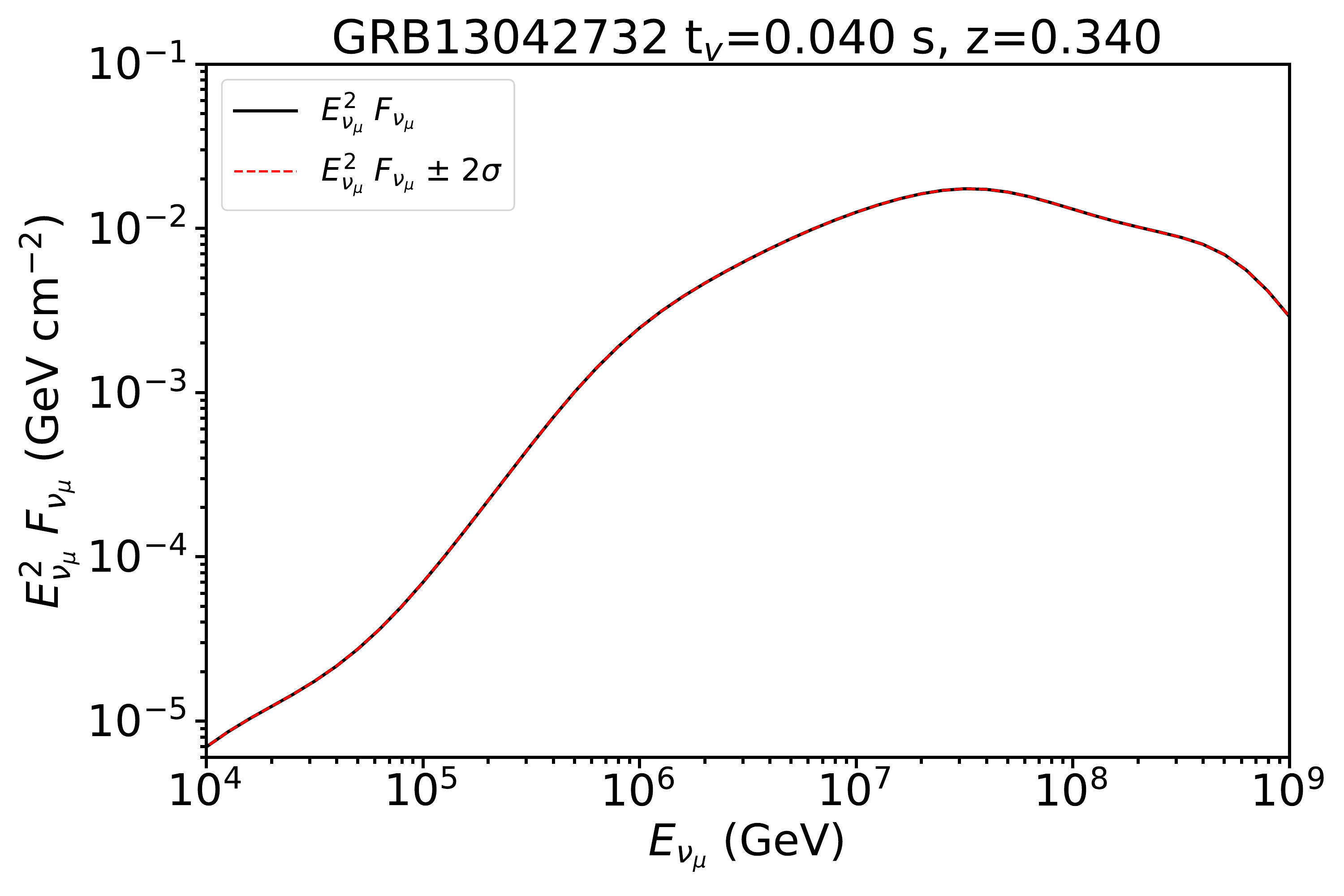}}
\caption{Expected neutrino fluence $\mathrm{E^2_{\nu_{\mu}}}\mathrm{F_{\nu_{\mu}}}$ as a function of the neutrino energy E$_{\nu_{\mu}}$. The $z$ and $t_{\rm v}$ values of each GRB are indicated in the figure, when known: (a) GRB08021273; (b) GRB14102845~\citep{xu}; (c) GRB08102853~\citepalias{tv2}; (d) GRB13042732~\citep{levan,tv3}. The grey thin lines indicate the results of 1000 simulations performed with the several randomly extracted values of $z$ and $t_{\rm v}$, when at least one of such parameters is unknown. The black thick line shows the mean of all the simulations or, when both $z$ and $t_{\rm v}$ are known, the resulting neutrino fluence. The red dashed lines delineate the error band around the neutrino fluence. In case both the minimum variability timescale and redshift are fixed, as for the GRB shown in (d), the error is very small; in this particular case, for example, it is estimated to be $\sim$ 3 \% around the neutrino fluence.}
\label{second}
\end{figure}

In this Appendix, the uncertainty due to missing parameters on individual GRB-neutrino fluences is explored, as explained in Sec.~\ref{sec:model}, and few examples of neutrino spectra obtained with NeuCosmA are reported. The unknown parameters investigated here are the redshift $z$ and the minimum variability timescale $t_{\rm v}$ only, as the uncertainty on the bulk Lorentz factor is marginalised by assuming $\Gamma$ values satisfying the correlation given in Eq.~\eqref{eq:correlation_lu}. The goal is hence to derive which among these parameters most affects the neutrino flux computation. \\
Several cases are shown, covering all the parameter combinations realized in the selected GRB sample, namely i) GRB08021273, a source with both $z$ and $t_{\mathrm{v}}$ unknown (Fig.~\ref{second}\ref{fig:both_unknown}); ii) GRB14102845, a source with measured $z=2.332$ but $t_{\mathrm{v}}$ unknown (Fig.~\ref{second}\ref{fig:tv_unknown}); iii) GRB08102853, a source with $z$ unknown and $t_{\mathrm{v}}=0.35$~s measured (Fig.~\ref{second}\ref{fig:z_unknown}); iv) GRB13042732 (also known as GRB130427A), the brigthest ever detected GRB in gamma rays, for which both $z=0.34$ and $t_{\mathrm{v}}=0.04$~s are measured (Fig.~\ref{second}\ref{fig:both_known}).\\
For each of these GRBs, 1000 simulations are performed extracting the unknown value of the missing parameter, either the redshift and/or the variability time, from a distribution of the same parameter as obtained from other known GRBs. From these examples, it follows that the minimum variability timescale contributes to the uncertainty on the neutrino fluence expected from GRBs significantly more than redshift. 
In fact, by comparing the cases (ii) and (iii) in Fig.~\ref{second}\ref{fig:tv_unknown} and Fig.~\ref{second}\ref{fig:z_unknown}, respectively, it is possible to note that the uncertainty due to the unknown value of $z$ is contained within $\sim$1 order of magnitude with respect to the mean flux, while it spans over several orders of magnitude when $t_{\rm v}$ is unknown. On the other hand, when both $z$ and $t_{\mathrm{v}}$ are measured, the error band on the neutrino flux is extremely reduced, as it is only due to the uncertainty in the measurements of spectral parameters. In these cases, it is not possible to distinguish the upper and lower bounds on the neutrino fluence from the mean fluence: an example is shown in Fig.~\ref{second}\ref{fig:both_known} for GRB13042732.\\
So far, the uncertainty related to the knowledge on $\Gamma$ was not considered, as justified by the assumption of a correlation that allows to infer its value, once the isotropic gamma-ray luminosity of the burst is given. The effects related to considering a different correlation are investigated in Appendix~\ref{appendixA}.

\section{Evaluating systematics on neutrino fluxes}
\label{appendixA}

\begin{figure*}
\subfigure[\label{fig:comparison}]{\includegraphics[width=0.55\columnwidth]{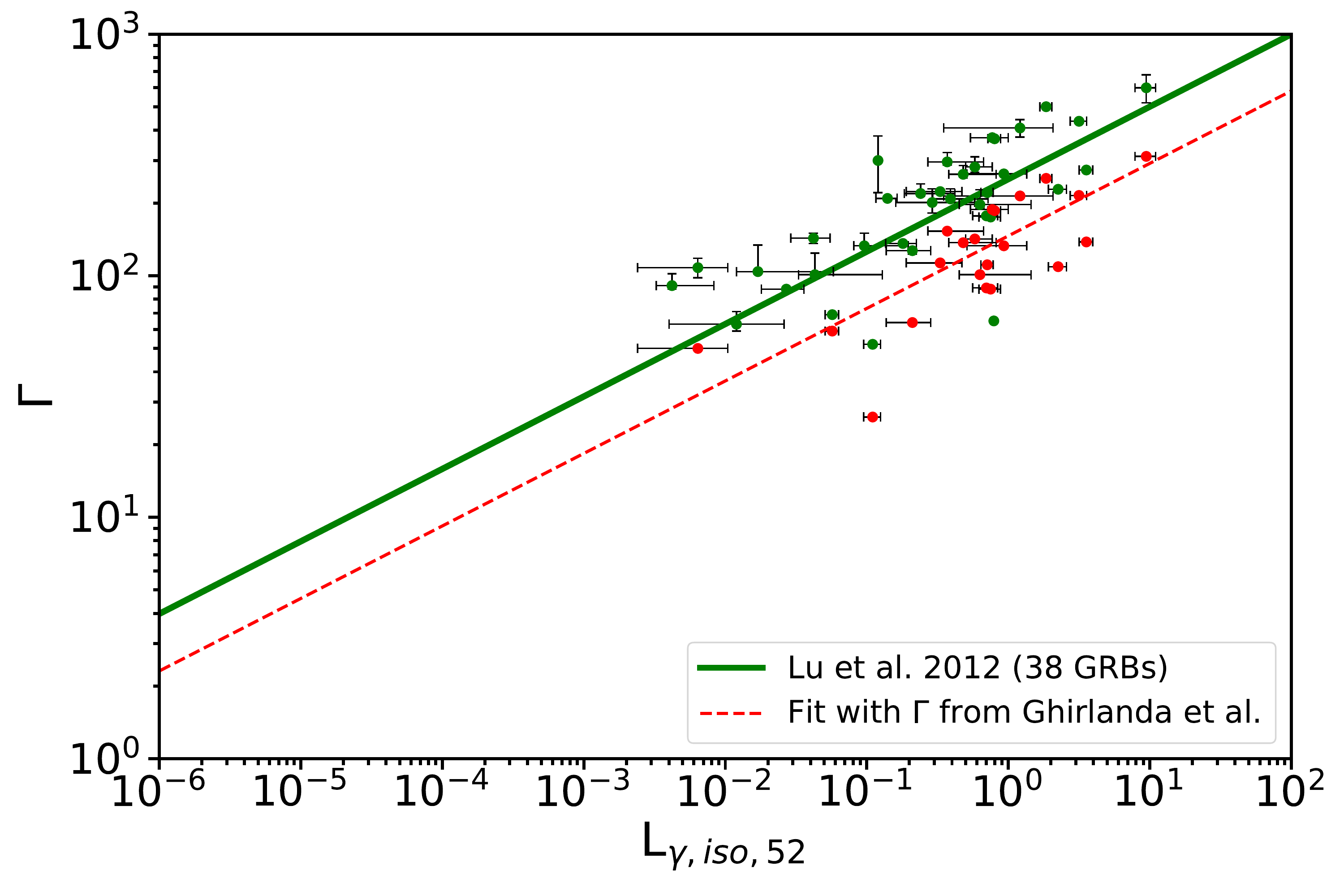}}
\subfigure[\label{fig:comparison_stacking}]{\includegraphics[width=0.55\columnwidth]{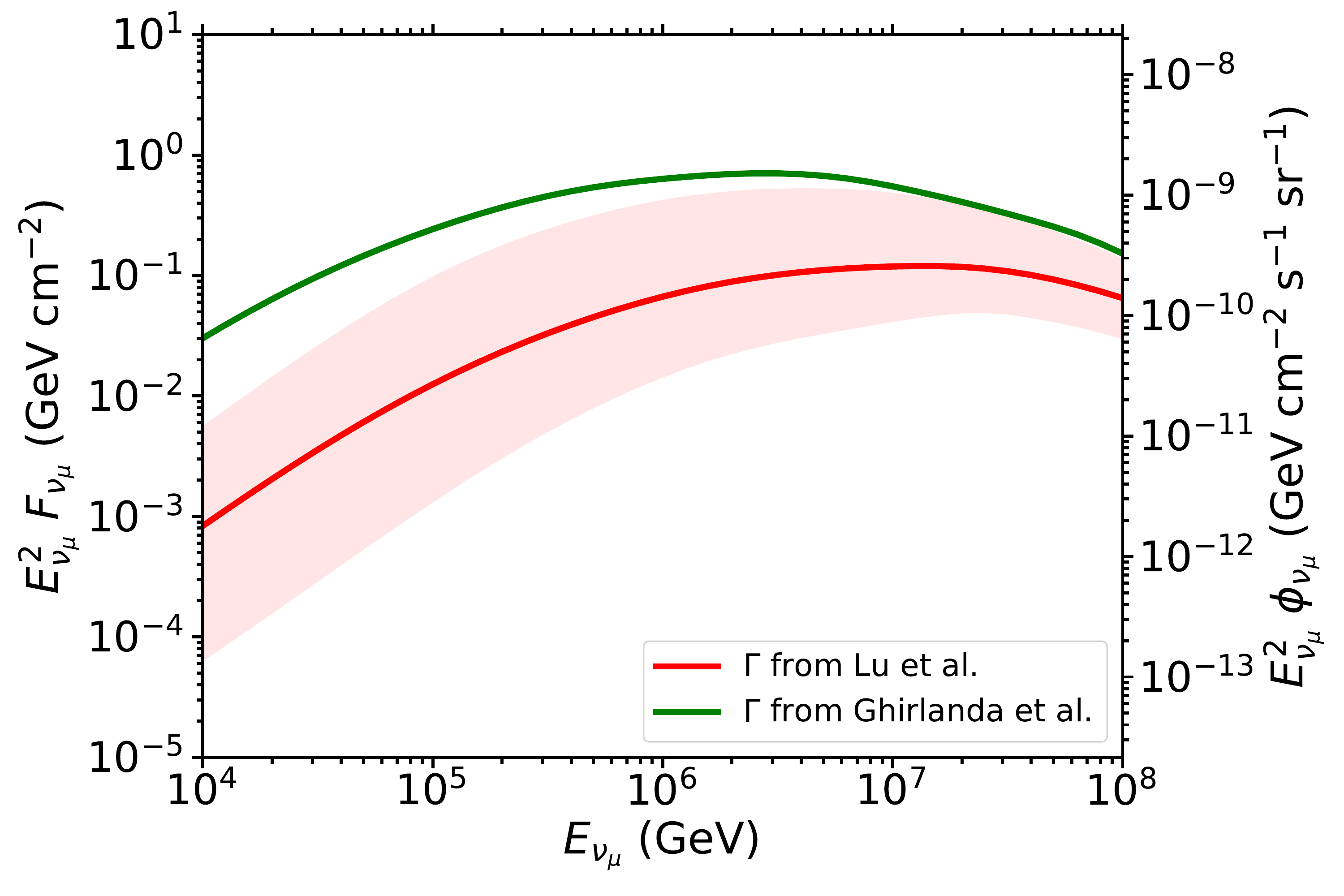}}
\caption{(a) The bulk Lorentz factor $\Gamma$ as a function of the isotropic equivalent gamma-ray luminosity $\rm L_{\gamma,iso}$. The green points represents GRBs in the sample studied by \citetalias{lu}. The red points, instead, are a subsample of the \citet{ghirlanda_corr} sample, containing only those GRBs in common with \citetalias{lu}, such that the values of $\Gamma$ come from \citet{ghirlanda_corr}, while the corresponding values of $\rm L_{\gamma,iso}$ are from \citetalias{lu}. The green solid and dashed red lines represent the best fits of each sample. (b) Total neutrino fluence $\mathrm{E^2_{\nu_{\mu}}}\mathrm{F_{\nu_{\mu}}}$ expected from the 784 GRBs in the ANTARES  2007-2017 sample (left-hand axis) and corresponding quasi-diffuse neutrino flux $\mathrm{E^2_{\nu_{\mu}}}\mathrm{\phi_{\nu_{\mu}}}$ (right-hand axis). The red and green lines show the different results obtained by either assuming a $\Gamma$-distribution according to \citetalias{lu} (see Eq.~\eqref{eq:correlation_lu}) or according to \citet{ghirlanda_corr} (see Eq.~\eqref{eq:ghirlanda}), respectively. The red shaded region indicates the error band around the stacking flux expected from \citetalias{lu}, as estimated in Sec.~\ref{cumulative}.}
\label{fig:ghirlanda_comparison}
\end{figure*}In addition to the parameter uncertainties considered so far, namely those due to the poor knowledge of redshift and minimum variability timescale (see Sec.~\ref{uncertainties} and Appendix~\ref{appendixC}), a further major source of uncertainty is related to the systematics on the treatment of the Lorentz factor, which could significantly affect the neutrino expectation from GRBs \citep{he2012}. In fact, the present analysis relies upon the correlation between the isotropic gamma-ray luminosity $\rm L_{\gamma,iso}$ and $\Gamma$ as derived by \citetalias{lu}, that has allowed the values of bulk Lorentz factor for each GRB in the sample to be determined by using Eq.~\eqref{eq:correlation_lu}, as explained in details in Appendix~\ref{appendixB}.\\
In order to evaluate the impact of such a method on neutrino expectations, the correlation found by \citet{ghirlanda_corr} was also tested. The latter one actually relates $\Gamma$ to the peak gamma-ray luminosity $\rm L_{\gamma,peak}$. Hence, as an intermediate step, the \citet{ghirlanda_corr} data sample was re-analyzed, to obtain the corresponding relation between $\Gamma$ and isotropic gamma-ray luminosity $\rm L_{\gamma,iso}$, similarly to the Eq.~\eqref{eq:correlation_lu}. Only common GRBs with respect to \citetalias{lu} were selected from the \citet{ghirlanda_corr} GRB sample, in order to consider the $\Gamma$ estimation from \citet{ghirlanda_corr} and the corresponding $\rm L_{\gamma,iso}$ from \citetalias{lu}. From this sample, the following correlation was found:
\begin{equation}
    \rm \Gamma_{G} \simeq 146 L_{\gamma,iso,52}^{0.30}.
    \label{eq:ghirlanda}
\end{equation}
The comparison among such a correlation and the one obtained by \citetalias{lu} is shown in Fig.~\ref{fig:ghirlanda_comparison}\ref{fig:comparison}. As visible, the Lorentz factor values obtained by \citet{ghirlanda_corr} are systematically lower by a factor $\sim$2 with respect to the values obtained by \citetalias{lu}. To quantify the impact of considering a reduced Lorentz factor on the expected number of neutrino events, the same method described in Sec.~\ref{uncertainties} was applied to the computation of neutrino spectra, namely for each GRB in the sample 1000 spectral simulations were performed with NeuCosmA, by extracting $\Gamma$ according to Eq.~\eqref{eq:ghirlanda}. By summing over all 784 GRBs, a revised stacking flux was obtained, as shown in Fig.~\ref{fig:ghirlanda_comparison}\ref{fig:comparison_stacking}. The spectral normalization appears now significantly higher with respect to the scenario described in Sec.~\ref{cumulative}, while the peak energy of the neutrino spectrum is shifted towards lower energies. \\\\With this novel neutrino spectrum, it is possible to re-run the data analysis chain, by optimising the track-quality cut $\Lambda_{\rm cut}$ consistently with the procedure described in Sec.~\ref{sec:optimisation}. Interestingly, the resulting cuts are found unaffected for most of the GRB sample. Nonetheless, the increased neutrino flux derived by adopting the \citet{ghirlanda_corr} implies a higher number of expected events in ANTARES with respect to the computation derived in Sec.~\ref{sec:results} for the correlation by \citetalias{lu}. In particular, this is estimated to be $\rm n_s \simeq 0.36$,
which is more than a factor 10 above the estimate presented in Eq.~\eqref{eq:number}. From the comparison with the estimated uncertainty due to missing information on redshift and variability timescale, which is contained within a factor of $\sim 5$ ($2\sigma$), it is possible to conclude that the leading source of uncertainty in neutrino spectral modeling is represented by the indirect knowledge of the bulk Lorentz factor of GRB jets. This conclusion is also supported by recent studies from \citet{he2012}.

\end{document}